# The Soil Reference Shrinkage Curve

V.Y. Chertkov*

*Agricultural Engineering Division, Faculty of Civil and Environmental Engineering, Technion, Haifa 32000, Israel*

**Abstract:** A recently proposed model showed how a clay shrinkage curve is transformed to the soil shrinkage curve at the soil clay content higher than a critical one. The objective of the present work was to generalize this model to the soil clay content lower a critical one. I investigated (i) the reference shrinkage curve, that is, one without cracks; (ii) the superficial layer of aggregates, with changed pore structure compared with the intraaggregate matrix; and (iii) soils with sufficiently low clay content where there are large pores inside the intraaggregate clay (so-called lacunar pores). The methodology is based on detail accounting for different contributions to the soil volume and water content during shrinkage. The key point is the calculation of the lacunar pore volume variance at shrinkage. The reference shrinkage curve is determined by eight physical soil parameters: (1) oven-dried specific volume; (2) maximum swelling water content; (3) mean solid density; (4) soil clay content; (5) oven-dried structural porosity; (6) the ratio of aggregate solid mass to solid mass of intraaggregate matrix; (7) the lacunar factor that characterizes the rate of the lacunar pore volume change with water content; and (8) oven-dried lacunar pore volume. The model was validated using available data. The model predicted value of the slope of the reference shrinkage curve in the basic shrinkage area is equal to unity minus the lacunar factor value, and is between unity and zero in the agreement with observations.

## 1. INTRODUCTION

The shrinkage curve of an aggregated soil, in general, is non-single valued because the volume of (macro) cracks between aggregates depends on sampling, sample preparation, sample size, and drying regime [1-6]. At the same time, the qualitative view of the soil shrinkage curve is kept at any interaggregate cracking (Fig.**1**). For this reason, to consider the effects of soil structure on shrinkage curve, one can introduce the *reference shrinkage curve* that, by definition, corresponds to shrinkage without interaggregate cracking and can be predicted in a single valued manner [7]. The term "reference" means that any measured shrinkage curve of a soil should go at least a little higher than its reference shrinkage curve at the expense of the crack volume contribution. Chertkov [7] quantitatively considered how a clay shrinkage curve is transformed into a corresponding soil reference shrinkage curve for soils with such high clay content, $c>c^*$ (the $c^*$ value gives a critical clay content) that lacunar pores (micro-cracks) [8] inside the intraaggregate clay lack (Fig.**2a**). This consideration was based on three assumptions [7].

*Assumption 1*. At given clay type and porosity $p$ of silt and sand grains coming into contact, there is such a critical clay content, $c^*$ that at a soil clay content $c>c^*$ lacunar pores inside clay and grain contacts lack (Fig.**2a**), but at a soil clay content $c<c^*$ lacunar pores exist and grain contacts can exist, at least at sufficiently small water contents (Fig.**2b**). The critical value $c^*$ to be defined by this assumption is [7]

$$c_* = [1 + (1/p - 1) \, v_z/v_s]^{-1} \tag{1}$$

where $v_s$ and $v_z$ are the relative volume of the clay solids and the relative oven-dried clay volume, respectively (the $v_s$ and $v_z$ values [9, 10] are connected with clay type).

*Assumption 2*. The shrinkage of an intraaggregate clay, aggregates, and soil, as a whole, starts simultaneously at total soil water content, $W_h$ corresponding to the point of maximum soil swelling (Fig.**1b**).

*Assumption 3*. Water in the clay pores of a thin rigid superficial layer of aggregates (Fig.**2a**) with modified pore structure (interface layer) and the layer volume, determine a soil reference shrinkage curve in the structural shrinkage area.

The soils with clay content $c>c^*$ only show a part of the observed peculiarities of the soil shrinkage curve compared with the clay shrinkage curve [7] (Fig.**1**). The objective of this work is to quantitatively consider the transition of a clay shrinkage curve to the soil reference shrinkage curve at lower than critical soil clay content, $c<c^*$. Notation is summarized at the end of the paper.

---
*Address correspondence to this author at the Agricultural Engineering Division, Faculty of Civil and Environmental Engineering, Technion, Haifa 32000, Israel; E-mail: agvictor@tx.technion.ac.il

## 2. THEORY

### 2.1. Preliminary Remarks

***Specification of the Intraaggregate Water Content Range***

As for soils with clay content $c>c^*$ [7], in the construction of a reference shrinkage curve at $c<c^*$ one should be interested in soil shrinkage starting from the water content $W_h$ of maximum swelling (Fig.**1b**). However, for quantitative determination of the reference shrinkage curve of a soil, the total range of the possible water content $w$ of the intraaggregate matrix, $0<w<w_M$ is important, but not only $0<w<w_h$ ($w_h<w_M$; $w_h$ corresponds to $W_h$). This total range corresponds to the matrix solid state [7] that extends up to the passage to a liquid state.

***Specification of the Intraaggregate Structure to Be Considered***

Intraaggregate configuration of contacting grains at $c<c^*$ (Fig.**2b**) is possible, but non-stable. Because of the non-homogeneous clay distribution in space (as a result of different non-homogeneous soil deformations) many (maybe the majority of) grains typically do not come into contact with neighbors and are surrounded by clay even at $c<c^*$ in the oven-dried state (Fig.**2c**). The principal difference between higher (Fig.**2a**) and lower (Fig.**2b** and **2c**) than critical soil clay content cases is not the grain contact, but the existence of lacunar pores at $c<c^*$. With that, the critical value, $c^*$ (Eq.(1)) is the same in the contact (Fig.**2b**) and general non-contact (Fig.**2c**) cases and is determined from the boundary (contact) situation in Fig.2b (see [7]). Indeed, if lacunar pores exist at a clay content $c<c^*$ in the contact case (Fig.**2b**), the more they should be at the same clay content in the non-contact case (Fig.**2c**). Thus, let us consider the general non-homogeneous clay-grain-lacunar pore configuration at $c<c^*$ (Fig.**2c**) with non-totally contacting grains even in the oven-dried state and will not use a specific relation that is inherent to the contact configuration in Fig.**2b** (see Results and Discussion).

***Specifics of Soils to Be Considered***

The specifics of soils with the more complex intraaggregate matrix at $c<c^*$ (Fig.**2c**), unlike that in Fig.**2a**, is connected with lacunar pores that usually have a small connectivity inside the clay matrix [8]. One should note three aspects flowing out of the sizes, existence, and volume evolution of the lacunar pores. First, the lacunar pores (Fig.**2b** and **2c**) can reach the sizes of intergrain spaces and are usually large compared to clay pores. For this reason they should be emptied (if water-filled) before the shrinkage starts (similar to structural pores [7] and after them). Second, the existence of lacunar pores (Fig.**2b** and **2c**) influences the specific volume $U$ of the intraaggregate matrix (see below) and the expression for the clay porosity $\Pi$ of the interface layer (see below). The third aspect relates to the direction of the volume evolution of lacunar pores at shrinkage in the total water content range, $0<w<w_M$ of the intraaggregate matrix. The intraaggregate clay (Fig.**2b** and **2c**) has surface parts coinciding with the interface layer surface and with the lacunar pore surface. The shrinkage of the intraaggregate clay leads to contraction of all clay surfaces. As a result the intraaggregate clay shrinkage should lead not only to aggregate shrinkage as a whole, but also to a lacunar pore volume increase inside the intraaggregate matrix (Fig.**2b** and **2c**). Thus, for soils with clay content $c<c^*$, along with the three above assumptions (see Introduction), one can make the additional Assumption 4 that the lacunar pore volume increases at soil shrinkage.

### 2.2. Summary of Features of the Reference Shrinkage Curve Model at Higher than Critical Soil Clay Content [7] that are Kept at Any Soil Clay Content

The assumptions (see Introduction) and part of the basic features of the theory for $c>c^*$ [7] are also applicable at $c<c^*$. Before developing the reference shrinkage curve model, specific for clay contents, $c<c^*$, it is worth to give a brief summary of those features noted at $c>c^*$ [7] that are general for any clay content and that will be used below.

The interface layer and intraaggregate matrix (Fig.**2**) give contributions $\omega$ and $w'$, respectively, to the total water content, $W$ of a soil as

$$W=w'+\omega, \qquad 0\leq W\leq W_h, \qquad 0\leq w'\leq w_h', \qquad 0\leq\omega\leq\omega_h=W_h-w_h' \qquad (2)$$

and contributions $U_i$ and $U'$, respectively, to the specific volume of aggregates, $U_a$ as

$$U_a=U'+U_i . \qquad (3)$$

The specific volumes of soil, $Y$, aggregates, $U_a$, and structural pores, $U_s$ are linked as

$$Y = U_a + U_s = U' + U_i + U_s \ . \tag{4}$$

$\omega$, $W$, $U'$, $U_a$, and $Y$ depend on the water contribution of the intraaggregate matrix, $w'$ in the range $0 \leq w \leq w_h'$ where $w_h'$ corresponds to the state of maximum soil swelling, $W_h$ [7]. To find the reference shrinkage curve $Y(W)$ one should find the $U'(w')$ and $\omega(w')$ functions as well as the constant specific volumes $U_i$ and $U_s$.

The $U'(w')$ dependence is simply connected with the shrinkage curve of the intraaggregate matrix in itself, $U(w)$. Unlike $U'$ and $w'$, $U$ and $w$ are the specific volume and water content of the same intraaggregate matrix per unit mass of the oven-dried matrix itself (but not the soil as a whole; Fig.**2**). The water contents $w$ and $w'$ are connected as $w' = w/K$ and $U$ and $U'$ values as $U' = U/K$ where $K$ is the ratio of the aggregate solid mass to the solid mass of the intraaggregate matrix (i.e., the solid mass of aggregates without the interface layer, see Fig.**2**). Thus, the auxiliary curve, $U'(w')$ is expressed through the shrinkage curve $U(w)$ of the intraaggregate matrix and $K$ ratio as [7]

$$U'(w') = U(w'K)/K, \qquad 0 \leq w \leq w_h' \ . \tag{5}$$

One can write the water contribution of the interface layer, $\omega(w')$ as [7]

$$\omega(w') = \begin{cases} 0, & 0 \leq w' < w_s' \\ \rho_w U_i \Pi F_i(w'), & w_s' \leq w' < w_h', \end{cases} \tag{6}$$

where $\rho_w$ is the water density; $F_i(w')$ is the volume fraction of water-filled interface clay pores at a given water content contribution, $w'$ of the intraaggregate matrix to $W$; $w_s'$ corresponds to the end point of structural shrinkage (Fig.**1b**); $U_i$ and $\Pi$ were defined above. $F_i(w')$ exists in two variants, is defined by the pore-size distributions of interface clay and intraaggregate clay (Fig.2), and does not depend on clay content.

The features that are expressed by Eqs.(2)-(6) and by the concrete form of the $F_i(w')$ function [7], take place at any soil clay content. The relations existing between constant (for a given soil) parameters $U_i$, $U_s$, $K$, and $W_h$ (Eqs. (30), (32)-(34) from [7]) are also kept at any clay content and will be additionally discussed below. Unlike that, the shrinkage curve $U(w)$ of the intraaggregate matrix (Fig.**2**) and modified clay porosity $\Pi$ of the rigid interface layer (Fig.**2**) are specific for clay content ranges $c \geq c^*$ (Fig.**2a**; see [7]) and $c < c^*$ (Fig.**2b** and **2c**; see below). Thus, for the estimation of the reference shrinkage curve of the soil with clay content in the range $c < c^*$ one should first find the corresponding shrinkage curve of the intraaggregate matrix, $U(w)$ and the clay porosity of the rigid interface layer, $\Pi$.

## 2.3. The Shrinkage Curve of the Intraaggregate Matrix in Relative Coordinates at Lower than Critical Soil Clay Content

The existence and volume evolution of lacunar pores (see Assumption 4) at $c < c^*$ leads to the essential modification of the shrinkage curve of the intraaggregate matrix $U(w)$ compared to the $c \geq c^*$ case [7]. First, let us consider the shrinkage curve of the intraaggregate matrix in Fig.**2c** in the dimensionless form, $u(\zeta)$ because such a presentation is simpler and more convenient for obtaining quantitative results. However, the final results will be presented in customary coordinates, $U(w)$.

The $u$ coordinate is the ratio of the intraaggregate matrix volume (Fig.**2c**) to its maximum volume in the solid state (at the liquid limit). The $\zeta$ coordinate is the ratio of the gravimetric water content of the intraaggregate matrix to its maximum value in the solid state (the liquid limit). As noted above, the lacunar pores (Fig.**2c**) are emptied (if they were filled in) before the intraaggregate clay shrinkage starts. This means that the $\zeta$ value corresponding to the initial shrinkage point (the maximum swelling point) of the intraaggregate matrix (Fig.**2c**) coincides with $\zeta$ for the similar point of the clay that contributes to the intraaggregate matrix. For the clay this point is $\zeta = \zeta_h \cong 0.5$ [7]. Note that the corresponding gravimetric water contents for the intraaggregate matrix and clay ($w$ and $\overline{w}$, respectively) differ.

As in the case of $c \geq c^*$ [7] the following derivation of $u(\zeta)$ is based on the dimensionless shrinkage curve, $v(\zeta)$ of the clay contributing to the intraaggregate matrix ($v$ is the relative clay volume similar to $u$ for the intraaggregate matrix). The clay shrinkage curve, $v(\zeta)$ is presented as [9, 10]

$$v(\zeta) = \begin{cases} v_z, & 0 \leq \zeta \leq \zeta_z \\[2mm] v_z + \dfrac{(1-v_s)^2}{4(v_z-v_s)(1-F_z)}(\zeta - \zeta_z)^2, & \zeta_z < \zeta \leq \zeta_n \\[2mm] v_s + (1-v_s)\zeta, & \zeta_n < \zeta \leq 1 \end{cases} \qquad (7)$$

with $v_s$ and $v_z$ being the relative volume of the clay solids and the relative oven-dried clay volume, respectively;

$$\zeta_z = ((v_z - v_s)/(1-v_s))F_z \qquad \text{and} \qquad \zeta_n = ((v_z - v_s)/(1-v_s))(2 - F_z) \qquad (8)$$

are the shrinkage limit and air-entry point, respectively, of the clay matrix; and $F_z$ is the saturation degree at $\zeta = \zeta_z$. One can estimate $F_z$ from $v_z$ and $v_s$ values [10] (see also [7]), then $\zeta_z$ and $\zeta_n$ from Eq.(8), and thereby one can express $v(\zeta)$ (Eq.(7)) through $v_s$ and $v_z$ only. The qualitative view of $v(\zeta)$ (pure clay) coincides with the curve in Fig.1a after replacement of $V$ with $v$, and $\overline{w}$ with $\zeta$.

If $c < c^*$ (Fig.**2c**) the relative volume of the clay $v(\zeta)$ (Eq.(7)) contributing to the intraaggregate matrix, and relative volume of the matrix, $u(\zeta)$ are connected as

$$v(\zeta) = (u(\zeta) - u_S - u_{lp}(\zeta))/(1-u_S), \quad v_z = (u_z - u_S - u_{lpz})/(1-u_S), \quad \text{and} \quad v_s = (u_s - u_S)/(1-u_S) \qquad (9)$$

where $u_z$, $u_s$, and $u_S$ are the relative volume of the intraaggregate matrix in Fig.**2c** in the oven-dried state, the relative volume of the solid phase of the intraaggregate matrix (silt and sand grains and clay particles), and the similar relative volume of the non-clay solids, respectively. $u_{lp}(\zeta)$ is the relative volume of the lacunar pores as a function of the relative water content $\zeta$, i.e., the ratio of their volume to the maximum volume of the intraaggregate matrix in the solid state; and $u_{lpz} \equiv u_{lp}(0)$. Relations from Eq.(9) directly follow from definitions of all entering values if $u_{lp}(\zeta)|_{\zeta=1}=0$; that is, lacunar pores are lacking at the maximum water content of the intraaggregate matrix (at its liquid limit). One can assume that this condition is fulfilled for soils with $c < c^*$, but being cohesive, i.e., capable of forming aggregates and keeping sufficiently appreciable shrinkage. Hereafter I am only interested in such soils (see Results and Discussion).

Replacing $v(\zeta)$, $v_z$, and $v_s$ in Eq.(7) from Eq.(9) one can obtain the shrinkage curve $u(\zeta)$ in relative coordinates for the intraaggregate matrix in Fig.2c as

$$u(\zeta) = \begin{cases} u_z, & 0 \leq \zeta \leq \zeta_z \\[2mm] u_z + u_{lp}(\zeta) - u_{lpz} + \dfrac{(1-u_s)^2}{4(u_z - u_s - u_{lpz})(1-F_z)}(\zeta - \zeta_z)^2, & \zeta_z < \zeta \leq \zeta_n \\[2mm] u_s + u_{lp}(\zeta) + (1-u_s)\zeta, & \zeta_n < \zeta \leq 1 \end{cases} \qquad (10)$$

Note that the shrinkage corresponding to the natural swelling takes place in the range of $0 < \zeta \leq \zeta_h \cong 0.5$ [7]. If the lacunar pores are lacking, i.e. $u_{lp}(\zeta) = 0$ (and $u_{lpz} = 0$), Eq.(10) reduces to equations for $c \geq c^*$ that coincide with Eq.(7) and (8) after the replacement all of the $v$ symbols with those of $u$ [7]. With that, for any clay content $\zeta$, $\zeta_z$, $\zeta_n$, $\zeta_h$, and $F_z$ are the same for the clay and intraaggregate matrix, but $u \neq v$, $u_s \neq v_s$, $u_z \neq v_z$, $u_n \neq v_n$, and $u_h \neq v_h$. Note also that for any clay content the relative volume of the non-clay solid phase, $u_S$ falls out of the expressions for $u(\zeta)$ (in the case $c < c^*$ see Eq.(10)). Finally, note that in addition to the relative characteristics of the intraaggregate matrix in the case of $c \geq c^*$ [7] (Fig.**2a**), $u_s$, $u_z$, and $u_S$, in the case of $c < c^*$ (Fig.**2c**) the relative volume contributions of lacunar pores, $u_{lp}(\zeta)$ and $u_{lpz}$ are added in Eq.(10). First of all let us consider the $u_{lp}(\zeta)$ dependence entering Eq.(10).

## 2.4. General Equation for the Lacunar Pore Volume in Relative Coordinates

The relative volume $u(\zeta)$ of the intraaggregate matrix (Fig.**2b** and **2c**) includes the constant relative volume $u_s$ of solids as well as the relative volumes of lacunar pores $u_{lp}$ and clay pores $u_{cp}$ as

$$u(\zeta) = u_s + u_{lp}(\zeta) + u_{cp}(\zeta), \qquad 0 \leq \zeta \leq 1 \ . \tag{11}$$

Note, that in general the lacunar pores appear, in the shrinkage process, at a water content (of the intraaggregate matrix) $w_L < w_M$. That is, in general $u_{lp}(\zeta) > 0$ only at $0 \leq \zeta \leq \zeta_L \leq 1$ where $\zeta_L$ corresponds to gravimetric water content $w_L$ (see below). In particular, variants $\zeta_L < \zeta_h \leq 1$ and $\zeta_h \leq \zeta_L \leq 1$ are possible where $\zeta_h = 0.5$ corresponds to the maximum swelling point $w_h$ of the intraaggregate matrix [7].

The small variation $du_{cp}$ of the clay pore volume at shrinkage ($du_{cp} < 0$), corresponding to a water loss $d\zeta < 0$, initiates both the variation $du$ of the intraaggregate matrix volume as a whole ($du < 0$) and the variation $du_{lp}$ of the lacunar pore volume inside the intraaggregate matrix (according to Assumption 4 $du_{lp} > 0$; Fig.**2c**). With that, according to Eq.(11) the volume balance takes place as

$$du_{cp} = du - du_{lp} \ , \qquad 0 \leq \zeta \leq 1 \ . \tag{12}$$

Based on the balance relation one can present $du_{lp}$ and $du$ as

$$du_{lp} = -k \ du_{cp} \ , \qquad 0 \leq \zeta \leq 1 \tag{13}$$

$$u = (1-k) \ du_{cp} \ , \quad 0 \leq \zeta \leq 1 \ . \tag{14}$$

According to Eq.(13) $k$ is the fraction of the clay pore volume decrease ($-du_{cp}$) that is equal to the lacunar pore volume increase ($du_{lp}$) at shrinkage, in the case that lacunar pore exist (i.e., at $\zeta \leq \zeta_L$). Below, for brevity, let us term $k$ to be the lacunar factor. In general $k$ can have values in the range $0 \leq k \leq 1$, and can depend on a number of soil characteristics, but not soil water content (see below).

According to Eqs.(13) and (14) $du_{lp}$ and $du$ are interconnected. Below the relation for $du_{lp}$ is used in the following form:

$$du_{lp}/d\zeta = -k \ du_{cp}/d\zeta \ , \qquad 0 < \zeta \leq \zeta_L \ . \tag{15}$$

The relative volumes of clay pores in the intraaggregate matrix, $u_{cp}(\zeta)$ and in the clay contributing to it, $v_{cp}(\zeta)$ are connected as (cf. Eq.(9)) $v_{cp}(\zeta) = u_{cp}/(1-u_S)$. That is,

$$u_{cp}(\zeta) = v_{cp}(\zeta) \ (1-u_S) = (v(\zeta) - v_s) \ (1-u_S) \ , \tag{16}$$

where $v(\zeta)$ is given by Eqs.(7) and (8). Replacing $u_{cp}$ in Eq.(15) from Eq.(16) one has

$$du_{lp}/d\zeta = -k \ (1-u_S) \ dv/d\zeta \ , \qquad 0 < \zeta \leq \zeta_L \ . \tag{17}$$

Integrating Eq.(17) with a known oven-dried value $u_{lp}(\zeta=0) \equiv u_{lpz}$ gives the general equation for the lacunar pore volume, $u_{lp}(\zeta)$ as

$$u_{lp}(\zeta) = u_{lpz} - k \ (1-u_S) \ (v(\zeta) - v_z), \qquad 0 < \zeta \leq \zeta_L \tag{18}$$

where $v(\zeta)$ is given by Eqs.(7) and (8). Note that the general equation for the lacunar pore volume (and together with that $k$, $u_{lpz}$, and $\zeta_L$) only has meaning at $c < c^*$. At $c = c^*$ (and even more at $c > c^*$) Eq.(18) degenerates to identity 0=0 because of the lack of lacunar pores; that is, $u_{lp} = 0$ and $u_{lpz} = 0$, $\zeta_L = 0$, and according to Eq.(13) $k = 0$. The soil characteristics, $k$, $u_{lpz}$, $u_S$, and $\zeta_L$ are considered in the following subsections.

## 2.5. Lacunar Factor and Lacunar Pore Volume in the Oven-Dried State

If lacunar pores are lacking at the maximum swelling point, $\zeta=\zeta_h$ and only appear in the course of shrinkage at $\zeta=\zeta_L<\zeta_h$ (for $\zeta_L$ see text after Eq.(11)), according to Eq.(13) $k=0$ at $\zeta_L<\zeta<\zeta_h$ and $0<k\leq1$ at $0<\zeta<\zeta_L<\zeta_h$. If the lacunar pores already exist at the maximum swelling point, $\zeta=\zeta_h$ (that is $\zeta_h<\zeta_L<1$), according to Eq.(13) $0<k\leq1$ in the total shrinkage range, $0<\zeta<\zeta_h$. I am interested in the $k$ value of a soil when $k>0$ in both of the above cases ($\zeta_L<\zeta_h$- lacunar pores appear in the course of shrinkage, or $\zeta_h<\zeta_L$ - lacunar pores already exist at maximum swelling). In general, the lacunar factor $k$ that characterizes the rate of the lacunar pore volume change with water content (cf. Eq.(17)) can depend on (i) the oven-dried lacunar pore volume ($u_{lpz}$) that plays the part of a characteristic value of lacunar pore volume, (ii) water content ($\zeta$), (iii) clay content (c), (iv) clay type ($v_s$, $v_z$), and (v) silt-sand characteristics (grain-size and -shape distributions, grain mineralogy - all designated by one symbol $\chi$). According to the stated above one can write

$$k=k(\zeta, u_{lpz}, c, v_s, v_z, \chi) \quad . \tag{19}$$

The possible dependence $k(\zeta)$ is the most interesting for us. Instead of $k(\zeta)$ in Eq.(13) one can consider $k(r)$ where $r\equiv u_{lp}(\zeta)/u_{cp}(\zeta)$ if $u_{lp}<<u_{cp}$, and $r\equiv u_{cp}(\zeta)/u_{lp}(\zeta)$ if $u_{lp}>>u_{cp}$. The former case takes place if the soil clay content $c<c^*$, but sufficiently close to $c^*$ so that the lacunar pore volume is small (Fig.2c). The latter case takes place if the soil clay content, $c$ is essentially lower than $c^*$ so that the clay pore volume is small. In both cases one has a small parameter $r(\zeta)<<1$ at $0<\zeta<\zeta_L$. That is, $k(r)\equiv k(0)=const>0$ at $0<\zeta<\zeta_L$. Based on these considerations we assume that for soils with intermediate clay contents $c$ (when the lacunar and clay pore volumes are comparable, $u_{lp}\sim u_{cp}$) the lacunar factor, $k$ is also const$>0$ at $0<\zeta<\zeta_L$. This assumption will be justified by the available data (see below).

Note that the oven-dried lacunar pore volume, $u_{lpz}\rightarrow0$ at $c\rightarrow c^*$ (with that, $k\rightarrow0$) and $u_{lpz}\rightarrow p$ (porosity of contacting grains) at $c\rightarrow0$ (with that, shrinkage disappears, $k\rightarrow1$, $u_z\rightarrow1$, and $u_S\rightarrow1-p$), but for a soil with $0<c<c^*$ $u_{lpz}$ in Eq.(19) is independent of $c$ and can have different values between a minimum and maximum.

Currently, dependence of $k$ from Eq.(19) is not available (except for $k(\zeta)=const$). For this reason one should consider $k$ and $u_{lpz}$ hereafter as independent fundamental physical properties of aggregated soils containing lacunar pores; that is, with clay content $0<c<c^*$. Finally, note that at any clay content in the range $0<c<c^*$ the constant $k>0$ at $0\leq\zeta<\zeta_L$ differs from $k=0$ at $\zeta_L<\zeta\leq1$ by a finite value. This is natural as $k$ should not be continuous and smooth at $\zeta=\zeta_L$, because with lacunar pores appearing at $\zeta=\zeta_L$, the state of the intraaggregate matrix changes qualitatively (similar to a phase transition).

## 2.6. The Relative Solids Volume of the Intraaggregate Matrix

Adding to Eq.(9) (for $v_s$) the relation

$$u_S/u_s=1-c \tag{20}$$

that flows out (at any clay content) of the above definition of $u_S$ and $u_s$, one can present $u_s$ and $u_S$ through clay type ($v_z$, $v_s$) and clay content (c) as

$$u_s=v_s[c+v_s(1-c)]^{-1} \tag{21}$$

$$u_S=v_s(1-c)[c+v_s(1-c)]^{-1} \quad . \tag{22}$$

## 2.7. The Relative Water Content of Lacunar Pore Appearance at Shrinkage

The lacunar pores (Fig.**2c**) appear in the intraaggregate matrix with shrinkage at some $\zeta=\zeta_L<1$. That is, $u_{lp}(\zeta_L)\equiv u_{lpL}=0$. After their appearance the lacunar pores grow in size up to $\zeta=0$. Taking $\zeta=\zeta_L<1$, $v(\zeta_L)\equiv v_L<1$, and $u_{lp}(\zeta_L)=0$ in the general equation for the lacunar pore volume (Eq.(18)), one comes to the equation

$$u_{lpz}-k\,(1-u_S)(v_L-v_z)=0 \quad . \tag{23}$$

Using Eq.(23) one finds the clay volume $v_L$ at the lacunar pore appearance to be

$$v_{\text{L}}=v_z+u_{\text{lpz}}/[k\,(1-u_\text{S})]\,, \tag{24}$$

at soil lacunar pore characteristics, $u_{\text{lpz}}$ and $k$, and $u_\text{S}$ from Eq.(22) at soil clay content $c$. Then, the relative water content, $\zeta_\text{L}$ of the lacunar pore appearance at shrinkage is found to be the solution of the equation $v(\zeta_\text{L})=v_\text{L}$ with $v(\zeta)$ from Eqs.(7) and (8) and $v_\text{L}$ from Eq.(24). Lacunar pore volume in the range $\zeta_\text{L}\le\zeta\le1$, is determined to be $u_{\text{lp}}(\zeta)=0$, and in the range $0<\zeta<\zeta_\text{L}$ $u_{\text{lp}}(\zeta)$ is given by Eq.(18). If $\zeta_\text{L}>\zeta_\text{h}=0.5$ and $v_\text{L}>v_\text{h}=v(\zeta_\text{h})=0.5(1+v_\text{s})$ (see Eq.(7)) the lacunar pores already exist at maximum swelling ($\zeta=\zeta_\text{h}$) when the shrinkage starts. If $\zeta_\text{L}\le\zeta_\text{h}$ and $v_\text{L}\le v_\text{h}$ the lacunar pores appear in the course of shrinkage at $\zeta=\zeta_\text{L}$ and $v=v_\text{L}$.

## 2.8. The Shrinkage Curve of the Intraaggregate Matrix in Customary Coordinates and its Slope in the Basic Shrinkage Area

One can transit from relative coordinates ($\zeta$, $u$) to customary ones ($w$, $U$) (specific volume vs. gravimetric water content) by the usual way as [9, 10]

$$w=((1-u_\text{s})/u_\text{s})(\rho_\text{w}/\rho_\text{s})\zeta\,, \qquad U=u/(u_\text{s}\rho_\text{s}) \qquad 0\le w\le w_\text{h} \tag{25}$$

where $w_\text{h}=w(\zeta_\text{h}=0.5)$ [7], and $\rho_\text{w}$ and $\rho_\text{s}$ being the density of water and (mean) density of solids (silt and sand grains and clay particles), respectively. Note, that Eq.(25) can be used not only to express the specific volume $U$ of intraaggregate matrix through its relative volume, $u$, but also for transition from the relative volume of lacunar pores, $u_{\text{lp}}(\zeta)$ to corresponding specific volume, $U_{\text{lp}}(w)$. In particular, $U_{\text{lpz}}=u_{\text{lpz}}/(\rho_\text{s}u_\text{s})$ and $U_{\text{lph}}=U_{\text{lp}}(w_\text{h})=u_{\text{lph}}/(\rho_\text{s}u_\text{s})=u_{\text{lp}}(\zeta_\text{h}=0.5)/(\rho_\text{s}u_\text{s})$.

According to Eq.(25) for the soil intraaggregate matrix one can write $dU/dw=(du/d\zeta)/(\rho_\text{w}(1-u_\text{s}))$. Replacing here $du/d\zeta$ in the basic shrinkage area ($\zeta>\zeta_\text{n}$) from Eq.(10) as $du/d\zeta=du_\text{lp}/d\zeta+(1-u_\text{s})$ and $du_\text{lp}/d\zeta$ from Eqs.(18) and (7) as $du_\text{lp}/d\zeta=-k\,(1-u_\text{s})(1-v_\text{s})=-k\,(1-u_\text{s})$ (see Eq.(9) for $v_\text{s}$) one finally obtains

$$dU/dw=(1-k)/\rho_\text{w}, \qquad w_\text{n}<w<w_\text{h}\;. \tag{26}$$

Thus, for soils with $c<c_*$ (Fig.**2c**), the $dU/dw$ value relating to the basic shrinkage area (Eq.(26)) varies between zero and $1/\rho_\text{w}$, depending on $k$ value (0<$k$<1). Only in the absence of lacunar pores ($u_\text{lp}=0$) at any water content ($c>c_*$, Fig.**2a**) does $k$=0, and the shrinkage curve slope of the intraaggregate matrix in the basic shrinkage area (Eq.(26)) is reduced to $1/\rho_\text{w}$ (cf. [7]) similar to pure clay (Fig.**1a**). Figure 3 shows in detail the variants of $U(w)$ dependence.

Unlike the shrinkage curve $V(\overline{w})$ of a disaggregated clay (Fig.**1a**), the point $U_\text{h}=U(w_\text{h})$ of the shrinkage curve $U(w)$ of the intraaggregate matrix (Fig.3) is only situated on the true saturation line, $1/\rho_\text{s}+w/\rho_\text{w}$ if $k$=0 in the vicinity of the $w_\text{h}$ point (Fig.**3a** and **3b**). In addition to the case of clay content $c>c^*$ (Fig.**3a**) this is also true for clay contents $c<c^*$ (Fig.**3b**) if the lacunar pores are absent at the beginning of shrinkage, that is, $w_\text{L}\le w_\text{h}$. If $c<c^*$ and $k$=const>0 in the whole range $0\le w\le w_\text{h}$ (Fig.**3c**), lacunar pores already exist at $w=w_\text{h}$ (that is, $w_\text{L}>w_\text{h}$), their volume remains empty, and the true saturation line, $1/\rho_\text{s}+w/\rho_\text{w}$ is not reached at swelling. In this case the point $U_\text{h}=U(w_\text{h})$ of the shrinkage curve $U(w)$ (Fig.**3c**) is situated on a 1:1 line (with unit slope) that is a pseudo saturation line.

Both $dU/dw<1/\rho_\text{w}$ at $w_\text{n}<w<w_\text{h}$ and swelling only up to a pseudo saturation line (Fig.**3c**), are direct consequences of the existence of lacunar pores in the intraaggregate matrix (Fig.**2c**) and evolution of their volume [$U_\text{lp}(w)=u_\text{lp}(\zeta(w))/(\rho_\text{s}u_\text{s})$] (Eqs.(18) and (25)) during shrinkage. At high clay content, $c>c_*$ when there are no lacunar pores (Fig.**2a**), these features of $U(w)$ are absent (Fig.**3a**) [7].

## 2.9. Additional Modifications Owing to Lacunar Pore Existence

Similar to the case of $c>c_*$ [7] in the case of $c<c^*$ a rigid clay porosity $\Pi$ of the interface layer coincides with the clay porosity of the intraaggregate matrix at maximum swelling. In the presence of lacunar pores (Fig. **2c**) this gives

$$\Pi=1-(u_\text{s}+u_\text{lph})/u_\text{h} \tag{27}$$

where $u_\text{lph}\equiv u_\text{lp}(\zeta_\text{h})$ and $u_\text{h}\equiv u(\zeta_\text{h})$.

The relations between $U_\text{s}$, $K$, $U_\text{i}$, and $W_\text{h}$ for the intraaggregate matrix in Fig.**2a**, Fig.**2b**, and Fig.**2c** flow out of the same prerequisites and formally coincide (see Eqs.(30), (32)-(34) from [7]). The differences between the cases of Fig.**2a** and Fig.**2c** are connected with the lacunar pore effect on $u_s$, $u_z$ (Eq.(9)), and $\Pi$ (Eq.(27)). Through $u_z$, $u_s$, and $\Pi$ the effect indirectly influences $K$, $U_s$, $U_i$ and $W_h$.

## 2.10. The Reference Shrinkage Curve and its Slope in the Basic Shrinkage Area

Figure 4 shows the consecutive steps to transform the shrinkage curve of a clay, $V(\overline{w})$ (Fig.**1a**) to the reference shrinkage curve of a soil, $Y(W)$ (Fig.**1b**). (i) Transition from $V(\overline{w})$ (not shown in Fig.**4**) to the shrinkage curve of an intraaggregate matrix, $U(w)$ ($w=\overline{w}$ $c$) that was considered earlier for $c \geq c^*$ [7] and above for $c<c^*$. (ii) Transition from $U(w)$ to the contribution, $U'(w')$ of the intraaggregate matrix to the specific volume of aggregates, $U_a(w')$ (Eq.(5)). (iii) Transition from the auxiliary curve $U'(w')$ to the reference shrinkage curve of aggregates $U_a(w')$ (in coordinates $w'$ and $U_a$) (Eq.(3)). (iv) Transition from $U_a(w')$ to the reference shrinkage curve of the soil, $Y(w')$ (in coordinates $w'$ and $Y$) (Eq.(4)). (v) Changing the scale along the water content axis (Fig.**4**) as a result of the transformation of the intraaggregate matrix water contribution, $w'$ to the total water content, $W$, accounting for the two variants (cf. Fig.1b) of the interface layer water contribution, $\omega(w')$ (Eqs.(2) and (6)). (vi) Transition from $U_a(w')$ to the same shrinkage curve in customary coordinates, $U_a(W)$; the two variants correspond to the two variants of transformation $W=w'+\omega(w')$ (cf. Fig.**1b**). (vii) Transition from $Y(w')$ to the same shrinkage curve in customary coordinates $Y(W)$ (only one variant is shown in Fig.**4**).

Note that five general features of the reference shrinkage curve at $c>c^*$ [7] are retained for $c<c^*$ (Fig.**2c**). In this case the model predicts yet two general features that are connected with the slope value of the reference shrinkage curve in the basic shrinkage area and appearance of the pseudo saturation line (Fig.**4**).

According to $w=w'K$ and Eq.(5) $dU'/dw'=dU/dw$ ($0\leq w'\leq w_h'$; $0\leq w\leq w_h=Kw_h'$). Therefore, $U'(w')$ in the range $0\leq w'\leq w_h'$ (Fig.**4**) is similar by shape to $U(w)$ in the range $0\leq w\leq w_h$ (Fig.**4**; note that Fig.**4** shows $U(w)$ corresponding to Fig.**3c**, i.e., for $k>0$ in the total range $0\leq w\leq w_h$). In particular, the basic shrinkage area, $w_n\leq w\leq w_h$ (Fig.**4**) where $dU/dw=(1-k)/\rho_w$ (Eq.(26)) corresponds to the basic shrinkage area, $w_n'=w_n/K\leq w'\leq w_h'$ (Fig.**4**) where $dU'/dw'=(1-k)/\rho_w$. It is essential that the reference shrinkage curve of a soil (Fig.**4**, curve $Y(w')$) and aggregates (Fig.**4**, curve $U_a(w')$) are similar in shape to the $U'(w')$ curve (Fig.**4**; see Eqs.(3) and (4)). That is, $dY/dw'\cong dU_a/dw'\cong dU'/dw'$ at $0\leq w'\leq w_h'$. In particular, they have the similar slope, $(1-k)/\rho_w$ in the basic shrinkage area, $w_n'\leq w'\leq w_h'$. Similar to the case of $c>c^*$ [7], at $W\leq W_s$ and $w'\leq w_s'$ (Fig.**4**) $W=w'$, $U_a(W)=U_a(w')$, and $Y(W)=Y(w')$. Hence, the slope of the observed reference shrinkage curve, $dU_a/dW$ or $dY/dW$ in the observed basic shrinkage area, $W_n\leq W\leq W_s$ (Fig.**4**) is also given by $(1-k)/\rho_w$

$$dU_a/dW=dY/dW=(1-k)/\rho_w, \qquad W_n\leq W\leq W_s \ . \tag{28}$$

Thus, for the structure in Fig.**2c**, the model quantitatively links the reference shrinkage curve ($Y(W)$) slope in the basic shrinkage area (Fig.**4**) with the lacunar factor, $k$ of the intraaggregate matrix with any possible value between zero and unity. Note that the slope does not depend on the $K$ ratio. In particular, for the structure in Fig.**2a** at any possible $K>1$ the slope is equal to unity [7].

Finally, in connection with Fig.**3c** for $U(w)$, it is worth noting that depending on lacunar pore connectivity the $Y(W)$ curve (Fig.**4**) can start on a pseudo saturation line (at $W=W_m$ in Fig.**4**) or reach the true saturation line and have an additional horizontal section in the range $W_m\leq W\leq W_m^*$ (Fig.**4**) as a result of lacunar pore filling.

# 3. MATERIALS AND METHODS

## 3.1. Input Parameters and Reference Shrinkage Curve Prediction

To predict the reference shrinkage curve of the soils with clay content $c<c^*$ (Fig.**2c**) one needs eight soil parameters. Six of them are the same as in the $c>c^*$ case [7] (Fig.**2a**): the oven-dried specific volume, $Y_z$ (Fig.**4**); maximum swelling water content, $W_h$ (Fig.**4**); mean solid density, $\rho_s$; soil clay content, $c$; oven-dried structural porosity, $P_z$; and ratio of aggregate solid mass to solid mass of intra-aggregate matrix, $K$ (Fig.**2**). Two additional parameters characterize the lacunar pores of an intraaggregate matrix and are the lacunar factor, $k$ and the specific lacunar pore volume in the oven-dried state, $U_{lpz}$ ($=u_{lpz}/(u_s\rho_s)$). Let us assume that all eight parameters are known, and show how the reference shrinkage curve can be found from the model under consideration. As noted above (see the text following Eq.(27)) the relations between parameters $U_s$, $U_i$, $W_h$, and $K$, are retained at any clay content (Eqs.(30), (32)-(34) from [7]) as

$$U_s=[(u_z/u_s)/(\rho_s K)+U_i]P_z/(1-P_z) \ . \tag{29}$$

$$(Y_z-U_i-U_s)K=u_z/(u_s\rho_s) \ . \tag{30}$$

$$W_h=0.5((1-u_s)/u_s)(\rho_w/\rho_s) \ . \tag{31}$$

$$U_i=U_h(1-1/K) \qquad \text{with } U_h=u_h/(u_s\rho_s) \ . \tag{32}$$

However, unlike in the case of $c>c^*$ [7] (Fig.**2a**), at $c<c^*$ (Fig.**2c**) the relative volume of an intraaggregate matrix at maximum swelling, $u_h$ in Eq.(32) contains an additional lacunar pore contribution, $u_{lph}$, that, in general, >0, (see Eq.(10)) at $\zeta=\zeta_h=0.5$) as

$$u_h=0.5(1+u_s)+u_{lph} \ . \tag{33}$$

This contribution complicates finding the reference shrinkage curve compared with the $c>c^*$ case [7] (Fig.**2a**). Equation (18) at $\zeta=\zeta_h$ gives $u_{lph}$ as

$$u_{lph}=u_{lpz}-k\,(1-u_S)\,(v_h-v_z) \qquad \text{with } v_h=0.5(1+v_s) \ (\text{see Eq.(7) at } \zeta=\zeta_h) \ . \tag{34}$$

Using the six relations from Eqs.(29)-(34), plus two relations from Eq.(9) for $v_z$ and $v_s$, Eq.(20), and relation $U_{lpz}=u_{lpz}/(u_s\rho_s)$ (in all, ten simple independent relations) at given $Y_z$, $W_h$, $\rho_s$, $c$, $P_z$, $K$, $k$, and $U_{lpz}$, one can find ten unknown parameters of intraaggregate clay ($v_z$ and $v_s$), intraaggregate matrix as a whole ($u_z$, $u_s$, $u_S$, $u_h$, $u_{lpz}$, $u_{lph}$), interface layer ($U_i$), and structural pores ($U_s$). After that one successively finds $F_z$ through $v_z$ and $v_s$ [10, 7], $v(\zeta)$ (Eqs.(7) and (8)), $u_{lp}(\zeta)$ (Eq.(18)), $u(\zeta)$ (Eq.(10)), $U(w)$ (Eq.(25); Fig.**4**), $U''(w')$ (Eq.(5); Fig.**4**), $U_a(w')$ (Eq.(3); Fig.**4**), and $Y(w')$ (Eq.(4); Fig.**4**). After that $v(\zeta)$ and $\Pi$ (Eq.(27)) successively give $F_i(w')$ [7], $\omega(w')$ (Eq.(6)), and $W(w')$ (Eq.(2); Fig.**4**). Given $Y(w')$ and $W(w')$, one can plot $Y(W)$ (Fig.**4**). Furthermore, one can obtain $U_{lph}=u_{lph}/(u_s\rho_s)$, $U_h=u_h/(u_s\rho_s)$ (Fig.**4**), and $w_L=w(\zeta_L)$ (Eq.(25); for $\zeta_L$ see the text after Eq.(24)).

The above general prediction algorithm of the reference shrinkage curve can be used at any clay content and any clay type that is characterized by $v_s$ and $v_z$. The particular case of $c>c^*$ [7] (Fig.**2a**) corresponds to $k=0$ and $U_{lpz}=0$. The above algorithm, in particular, shows that the input parameters that were used ($Y_z$, $W_h$, $\rho_s$, $c$, $P_z$, $K$, $k$, and $U_{lpz}$) permit one to find both clay matrix parameters, $v_s$ and $v_z$, and intraaggregate matrix parameters, $u_s$ and $u_z$. However, $v_s$ and $v_z$ (unlike $u_s$ and $u_z$) can also be found independently of the above input parameters from measurements of the clay contributing to the soil [9, 10]. Thus, $v_s$ and $v_z$, if they are available, can be used as independent input parameters instead of $Y_z$ and $W_h$.

All of the above input parameters ($Y_z$, $W_h$, $\rho_s$, $c$, $P_z$, $K$, $k$, and $U_{lpz}$) have clear physical meaning and in general can be measured or estimated from measurements that are independent of an observed shrinkage curve. As applied to the three new parameters, $K$, $k$, and $U_{lpz}$ this issue is discussed in the Results and Discussion. However, unlike other parameters, currently corresponding data on $K$, $k$, and $U_{lpz}$ are unavailable. For this reason, to use the above algorithm, $K$, $k$, and $U_{lpz}$ are estimated below on the united basis as follows. Using the predicted and experimental values of the specific volume ($Y$ and $Y_e$) for the water contents participating in the shrinkage curve measurements, one can calculate the maximum relative difference $\delta=\max(|Y-Y_e|/Y_e)$ between the predicted and measured values. Then, one can consider $K$, $k$, and $U_{lpz}$ to be the fitted parameters that correspond to the minimum value of the maximum relative difference, $\delta$. It is noteworthy that, unlike usual fitting, all three found values of $K$, $k$, and $U_{lpz}$ can be directly compared with corresponding independent values that are immediately found using the measured shrinkage curve as $K=W_h/w_h'$ (Fig.**4**), $k=1-S$ ($S$ is a curve slope to be observed in the basic shrinkage area), and $U_{lpz}$ is independently found from the shear of the true saturation line relative to the pseudo saturation one (see below). The coincidences of different estimates of $K$, $k$, and $U_{lpz}$ are not a trivial fact. They are not guaranteed in the case of non-adequate modeling.

### 3.2. Data Used

Available data that could be used for checking the model are very limited. Full data sets, including both shrinkage curve data and input parameters, are currently unavailable. The most suitable data sets are in [11]. Using the model we analyzed seven experimental shrinkage curves from Fig.**3** of [11] for a ferruginous soil (from the Congo) of A, B1, B2, and AB horizons as well as for a ferralitic soil (from the Congo) of A, B1, and B2 horizons (the data of the ferralitic soil horizon AB with $c>c^*$ has been considered in [7]). These data are reproduced in Fig.**5-11** (white squares), respectively. They were obtained by continuous monitoring of soil sample shrinkage [5].

The solid density, $\rho_s$ for the data in Fig.**6-8**, **10**, and **11** was estimated (Table **1**) from the location of the true saturation line $Y=1/\rho_s+W/\rho_w$ in Fig.**3** of [11]. For the data in Fig.**5** and 9 [11] does not give the position of the saturation line. For the data in Fig.**5** $\rho_s$ was taken (Table **1**) to be similar to that for other ferruginous soil horizons (Fig.**6-8**) from [11]. For data in Fig.**9** $\rho_s$ was taken (Table **1**) to be similar to that for other ferralitic soil horizons (Fig. **10** and **11**) from [11]. Except for the $\rho_s$ estimates, the data from [11] on clay content, $c$ (Table **1**) were used. Input data on $Y_z$, $W_h$, $W_m$, $W_h^*$, and $W_m^*$ (Table **1**; see Fig.4) correspond to the experimental points in oven-dried state, ($Y_z$), maximum swelling ($W_h$), the intersection between the shrinkage curve and pseudo saturation line ($W_m$), and between the curve and true saturation line ($W_h^*$, $W_m^*$) in Fig.**5-11**. Input data on $P_z$ were estimated as $P_z=U_s/Y_z$ (for $U_s$ see also Table **2**) from the size, $W_m-W_h=U_s\rho_w$ (see Fig.4) of a horizontal section of shrinkage curve at water contents higher than maximum swelling, $W>W_h$ in Fig.**5-11**. Note that the oven-dried structural porosity, $P_z$ could be estimated from an aggregate-size distribution [12]. However, such data were unavailable for the above seven soils. In addition, based on the continuous monitoring data in Fig.**5-11** (white squares), I estimated $K=W_h/w_h'$ (Table **1**), $S$ (Table 1) where $S$ is the experimental shrinkage curve slope in the basic shrinkage area, and a shear ($W_h^*-W_h$) (or ($W_m^*-W_m$); see Fig.4) between the true and pseudo saturation lines, to compare with the values, $K_{fit}$, 1-$k_{fit}$, and $U_{lph}$ (the latter is found through $k_{fit}$ and $U_{lpzfit}$), respectively (see $K_{fit}$, $k_{fit}$, $U_{lpzfit}$ in Table **1**).

As one can see from Table **1** and Fig.**5-11** the seven data sets are interesting for checking the model because they show various features. The observed clay content and slope in the basic shrinkage area vary from curve to curve in the wide range ($0.065 \le c \le 0.648$; $0.071 \le S \le 0.879$). The mean solid density also essentially changes from 2.337 to 2.608 g cm$^{-3}$. The experimental curves include those with a horizontal section before the shrinkage starts (that is, $U_s>0$ and $P_z>0$) and without that (that is, $U_s=0$ and $P_z=0$). The experimental curves include those with both possible types of behavior in the structural shrinkage area (type 1 and type 2 in Fig.4). Finally, the experimental shrinkage curves include those that start at the pseudo saturation line (that is, $U_{lph}\rho_w=W_h^*-W_h>0$ or $U_{lph}\rho_w=W_m^*-W_m>0$) and those that start at the true saturation line (that is, $U_{lph}\rho_w=W_h^*-W_h=0$ or $U_{lph}\rho_w=W_m^*-W_m=0$).

## 4. RESULTS AND DISCUSSION

### 4.1. The Model Prediction Results and Comparison with Data

Figures **5-11** show the predicted shrinkage curves, $Y(W)$ (for soils in Fig.**5**, **8-10** $Y(W)=U_a(W)$ since $U_s=0$). The predicted auxiliary curves, $U(w)$, $U'(w')$, $U_a(w')$, and $U_a(W)$ if $U_s>0$ (see Fig.4) were omitted in order not to encumber the figures. Parameter calculation of a soil and the clay contributing to it, is a part of the reference shrinkage curve prediction. Tables **2** and **3** show the parameters for the seven soils.

One can compare the fitted values, $K_{fit}$ and $k_{fit}$, (Table 1) with independent estimates $K=W_h/w_h'$ and $k=1-S$ (Table **1**). Similarly, one can compare $U_{lph}$ (Table **2**), found from $k_{fit}$ and $U_{lpzfit}$ (Table **1**), with ($W_h^*-W_h)/\rho_w$ or ($W_m^*-W_m)/\rho_w$ (see Fig.4 and Table **1**). The estimates $K=W_h/w_h'$ and $k=1-S$ simply coincide with corresponding $K_{fit}$ and $k_{fit}$ with an accuracy higher than 0.001. $U_{lph}$ differs from ($W_h^*-W_h)/\rho_w$ or ($W_m^*-W_m)/\rho_w$ by less than 0.001. These coincidences give three independent arguments in favor of the feasibility of the model.

Table **1** shows the small $\delta$ values for the $K_{fit}$, $k_{fit}$, and $U_{lpzfit}$ values found. These $\delta$ are within the limits of the relative errors of the measurement method [5]. This is also evidence in favor of the feasibility of the model.

In addition, for each soil (Fig.**5-11**) the data on structural shrinkage confirm the model because they correspond to either Curve 1 or 2 in Fig.**4**. The type of shrinkage curve in the structural shrinkage area is determined by the pore-size distribution in the interface layer of aggregates. Note that the simplest possible pore-size distribution was used in [7]. The possible pore-size distribution including two and more modes [12] predicts that shrinkage curves with more than one inflection point in the structural shrinkage area are possible.

For data sets (Table 1) that correspond to Fig.5-9, $v_L > v_h$ (Table 3). This prediction is in agreement with $U_{lph} > 0$ (Table 2) and $W_h^* - W_h > 0$ (or $W_m^* - W_m > 0$; Table 1) for the soils. Similarly, if $v_L \leq v_h$ (Fig. 10 and 11; Table 3), $W_h^* - W_h = 0$ (or $W_m^* - W_m = 0$; Table 1) and $U_{lph} = 0$ (Table 2). This also speaks in favor of the model.

## 4.2. Checking the Predicted General Features of the Reference Shrinkage Curve

All of the above arguments in favor of model feasibility, including the comparison between predicted and measured shrinkage curves, simultaneously confirm the predicted general features of the reference shrinkage curve, the five that exist at any clay content and have been already noted at $c > c^*$ [7] and two new features that only reveal themselves at $c < c^*$ and are connected with the existence and volume evolution of lacunar pores in the intraaggregate matrix: (i) the reference shrinkage curve slope in the basic shrinkage area is numerically equal to unity minus the lacunar factor, $k$ and varies between zero and unity depending on the soil $k$ value (for which $k$ does not depend on water content); and (ii) the shear along the water content axis, $W_h^* - W_h$ (or $W_m^* - W_m$; see Fig.4) between the pseudo and true saturation lines, is stipulated by the volume $U_{lph}$ of the empty lacunar pores at the maximum swelling point, $W = W_h$ ($U_{lph} = (W_h^* - W_h)/\rho_w$ or $U_{lph} = (W_m^* - W_m)/\rho_w$). If $U_{lph} = 0$ (e.g., Fig.10 and 11; Table 2), these saturated lines coincide ($W_h^* = W_h$ and $W_m^* = W_m$).

## 4.3. Increasing the Interface Layer Thickness with Clay Content Decrease

One can see that for soils of a given type (e.g., ferralitic or ferruginous) the aggregate/intraaggregate mass ratio, $K$ ($K_{fit}$ in Table 1) and specific volume of the intraaggregate matrix, $U_i$ (Table 2) essentially increase as clay content decreases. To understand this result one should consider two opposite cases: $c \to 1$ and $c \to 0$. When clay content $c \to 1$ one deals with a nearly pure clay that as an intraaggregate matrix fills nearly all aggregate volume. That is, the thickness of the interface layer (Fig.2a) and its specific volume $U_i$ strive to zero, and the $K$ ratio (aggregate solid mass to intraaggregate that) strives to unity (cf. Eq.(32)). Thus, at $c \to 1$, $U_i \to 0$, and $K \to 1$, aggregates disappear because of the transition to a paste-like state of the intraaggregate matrix with a high clay content and without an interface layer (or a very thin one).

When the clay content $c \to 0$ one deals with aggregates that are nearly totally filled by contacting silt and sand grains and for this reason they are quite rigid. One can consider that such aggregates nearly totally consist of a rigid interface layer. That is $K \to \infty$ (because there is no intraaggregate matrix and its solid mass is zero), and $U_i \to U_h$ (see Eq.(32)). Thus, at $c \to 0$, $U_i \to U_h$, and $K \to \infty$, aggregates also disappear because they degenerate to silt and sand grains. It follows from the above considerations that with clay content decrease in the range $0 < c < 1$ the $K$ ratio increases in the range $1 < K < \infty$. At sufficiently high clay content $c^* < c < 1$ the $K$ ratio values are close to unity [7]. However, at $c < c^*$ the $K$ ratio can essentially exceed unity (see $K_{fit}$, Table 1), and the rigid interface layer cannot be considered to be a thin one as at $c > c^*$. Its thickness can make up an essential part of aggregate size.

## 4.4. Possible Ways to Obtain Three New Input Parameters as Given, but not Fitted

The general approach to estimate $K$ independently of a measured shrinkage curve was noted in [7]. $K$ is calculated through $U_i$ (Eq.(32)). In turn, the specific volume, $U_i$ of the interface layer (see Fig.2) is expressed through the layer thickness, oven-dried aggregate-size distribution, and soil texture. Then $K$, as an input parameter, is replaced with more than one elementary input parameter that is directly measured. This way was considered for the thin interface layer, at $c > c^*$ [13]. The interface layer of a finite thickness can be similarly considered.

To estimate the oven-dried lacunar pore volume $u_{lpz}$ ($U_{lpz} = u_{lpz}/(\rho_s u_s)$) independently of a measured shrinkage curve, one can use the simple relation between the dry aggregate bulk density $\rho_b$ and mean solid density of the soil, $\rho_s$ as

$$\rho_b = \rho_s u_s / [u_S + v_z(1 - u_S) + u_{lpz}] \qquad (35)$$

where $u_s$ and $u_S$ are expressed through $c$ and $v_s$ (Eqs.(21) and (22)). Thus, measuring $\rho_b$, $\rho_s$, and $c$ for the soil, and knowing $v_s$ and $v_z$ (from clay measurements or through other input parameters; see above) one can estimate $u_{lpz}$. Unfortunately, for the above seven data sets (Fig.**5-11**) data on the dry aggregate bulk density, $\rho_b$ were not available.

To estimate the lacunar factor $k$ independently of a measured shrinkage curve, one can use data on specific soil ($Y_h$) or aggregate ($U_{ah}$) volume at maximum swelling (Fig.**4**). Then, the specific volume of the intraaggregate matrix at maximum swelling $U_h=U_{ah}=Y_h-U_s$ (Fig.**4**). Knowing $U_h$ and $u_h=U_h\rho_s u_s$ (Eq.(25)) one then also knows $u_{lph}=u_h-0.5(1-u_s)$ (Eq.(10)) at $\zeta=\zeta_h=0.5$. Finally, given $u_{lpz}$ (see above) and $u_{lph}$, one can estimate $k$ from Eq.(18). This way was tried, and $k$ values were obtained that coincide with those indicated in Table 1 as $k_{fit}$. Still another way to estimate $k$ (in the future) is connected with the construction of an approximate explicit dependence $k(u_{lpz})$ based on Eq.(19).

## 4.5. Estimating the Porosity of Contacting Grains and Critical Soil Clay Content

To estimate the critical soil clay content, $c^*$(Eq.(1)) one needs the porosity $p$ of the contacting silt and sand grains contributing to the soil. This parameter is an independent soil property and cannot be exactly found using the model. However, the model enables us to estimate the range of possible $p$ values of the soil. In the special case of an intraaggregate structure when silt and sand grains come into contact, at least at $\zeta=0$ (Fig.**2b**), there is a specific relation as $u_z=u_S/(1-p)$ between the oven-dried volume ($u_z$), non-clay solids volume ($u_S$), and porosity of contacting grains ($p$). This relation immediately flows out of the definitions of $u_z$, $u_S$, and $p$. Based on this specific relation one can express $u_z$, $u_{lpz}$, and $k$ characteristics through the elementary soil parameters, $c$, $v_s$, $v_z$, and $p$ (which is beyond the scope of this work). In this work the general case is considered (Fig.**2c**) that includes the particular case of Fig.**2b**. However one can use the specific relation to estimate the upper boundary, $p_{max}$ of possible (for a soil) $p$ values. In the general case (Fig.**2c**) the relation turns into an inequality, $u_z>u_S/(1-p)$. That is,

$$p<p_{max}=1-u_S/u_z \ . \tag{36}$$

We are interested in soils with $c<c^*$. One can use this inequality with $c^*$ from Eq.[1] to estimate the lower boundary, $p_{min}$ of possible (for a soil) $p$ values as

$$p>p_{min}=\{1+(v_s/v_z)[(1-c)/c]\}^{-1} \ . \tag{37}$$

Table **3** shows boundaries of the range $p_{min}<p<p_{max}$ for the abovementioned seven soils. One can see that for the five soils with relatively high clay content (Fig.**6**, **7**, **9-11**) this range is quite narrow (*i.e.*, $(p_{max}-p_{min})/p_{min}<<1$) and the possible $p$ value itself is quite high ($p>0.5$). For the two soils with relatively low clay content (Fig.**5** and **8**) this range is not narrow ($(p_{max}-p_{min})/p_{min}\sim1$) and the possible $p$ value itself is relatively low. Thus, these model estimates of $p_{min}$ and $p_{max}$ suggest some link between the soil clay content and porosity of silt and sand grains when they come into contact, although such contacts are typically lacking in real soils even at small clay content. Table **3** also shows values of $p_{av}=(p_{min}+p_{max})/2$ and $c^*$ estimated for the seven soils from Eq.(1) at $p=p_{av}$.

## 4.6. The Possible Effect of a Difference between Densities of Clay Solids and Grains

In all equations of this work that include clay content $c$ (and $c^*$) as in [7], $c\equiv c_v$ being the clay particle volume fraction of all solids. However, available data on clay content $c$ (Table **1**) that are used (similar to the clay content data that were used in [7]) are $c\equiv c_w$ being the clay particle weight fraction of all solids. That is, one considers that $c_v=c_w=c$. This means, as noted [7], that for simplicity one neglects a possible small difference $\Delta\rho$

$$\Delta\rho=\rho_{cl}-\rho_{ss} \tag{38}$$

between the densities of clay ($\rho_{cl}$) and grain ($\rho_{ss}$) solids and correspondingly, the difference between the clay content by weight, $c_w$ and the volume fraction of clay solids, $c_v$. The modification accounting for $\Delta\rho\neq0$ and $c_w\neq c_v$ is possible.

By definition of the mean solid density $\rho_s$ one can write

$$\rho_s=\rho_{cl} c_v+\rho_{ss}(1-c_v) \ . \tag{39}$$

According to Eq.(38) and (39)

$$\rho_{ss}=\rho_s-\Delta\rho \ c_v \qquad \text{and} \qquad \rho_{cl}=\rho_s+\Delta\rho \ (1-c_v) \qquad (40)$$

(at $\Delta\rho=0$ $\rho_{ss}=\rho_{cl}=\rho_s$). By definition of all entered values

$$c_w=\rho_{cl} \ c_v/\rho_s \ . \qquad (41)$$

Replacing in Eq.(41) $\rho_{cl}$ from Eq.(40) and introducing $\varepsilon\equiv\Delta\rho/\rho_s$ one can express $c_w$ through $c_v$ as

$$c_w=(1+\varepsilon)c_v-\varepsilon \ c_v^2 \qquad (42)$$

(at $\Delta\rho=0$ $\varepsilon=0$ and $c_w=c_v$). It follows from Eq.(42) the reverse link as

$$c_v=\{(1+\varepsilon)-[(1+\varepsilon)^2-4\varepsilon \ c_w]^{1/2}\}/(2\varepsilon) \qquad (43)$$

and a useful relation, $\varepsilon\equiv\Delta\rho/\rho_s=(c_w-c_v)/[c_v(1-c_v)]$. Based on Eq.(42) and (43) one can find different approximations at small $\varepsilon$, $c_v$, or $c_w$ as well as at $c_v$ or $c_w$ close to unity. Using Eq.(43) at additionally given $\Delta\rho$ (it is assumed that $\rho_s$ is known) one can pass from an experimental $c_w$ value (as $c$ values in Table **1**) to corresponding experimental $c_v$ values. Then one can use this experimental $c_v$ to check the model as above. Similarly, after calculating the $c_v^*$ value from Eq.(1) (as $c^*$ values in Table **3**) one can pass to critical clay content $c_w^*$ by weight using Eq.(42) (at given $\Delta\rho$ and known $\rho_s$).

Unfortunately, data on $\Delta\rho$ for the above seven soils were not available. For this reason the possible effect of $\Delta\rho$ was estimated giving its values "manually". According to its definition (Eq.(38)) $\Delta\rho>0$ and $\Delta\rho<0$ are possible. However, the $|\Delta\rho|$ value should be in the range $0<|\Delta\rho|<0.3-0.5$ g cm$^{-3}$ because clay and grain solids density is usually in the range $2.6<\rho_s<3$ g cm$^{-3}$, and $|\Delta\rho|<(3-2.6)\sim0.5$ g cm$^{-3}$. The main result of the calculations with $\Delta\rho=0$; 0.1; 0.3; 0.5 g cm$^{-3}$ is as follows. The effect of $\Delta\rho$ variation (in the indicated range) on the predicted shrinkage curves (Fig.**5-11**) is negligible. However, the $\Delta\rho$ variation can, to some extent, influence the predicted clay parameters ($v_s$, $v_z$; Table **3**) and critical clay content, both $c_v^*$ and $c_w^*$. In any case, for a possibly more accurate prediction, data on $\Delta\rho$ are desirable.

### 4.7. The Model Applicability Condition

According to its physical meaning the specific clay volume $v_L$ (Eq.(24)) corresponding to the appearance of lacunar pores at shrinkage, should be $\leq1$. Thus, the inequality

$$v_L\leq1 \qquad (44)$$

with $v_L$ from Eq.(24) at a given set of independent soil parameters $v_s$, $v_z$, $c$, $k$, and $u_{lpz}$ ($u_S$ in Eq.(24) is a function of $c$ and $v_s$ from Eq.(22)), is the model applicability condition. At sufficiently low soil clay content, $c$ depending on clay type ($v_s$, $v_z$) and lacunar pore properties ($k$ and $u_{lpz}$) the condition from Eq.(44) can be violated. This means that a soil with such a parameter set ($v_s$, $v_z$, $c$, $k$, and $u_{lpz}$) is a cohesionless one with some clay content, but without appreciable shrinkage. Note, that the model applicability condition is closely connected with the above condition, $u_{lp}(\zeta)_{\zeta=1}=0$ (see the text after Eq.(9)). Note also, that for all the above seven soils this condition is met (for $v_L$ see Table **3**) even at quite small clay content of Fig.**5**.

### 5. CONCLUSION

The qualitative differences between the observed shrinkage curves of aggregated soils and clays contributing to them suggest new features of the intraaggregate soil structure. An approach to quantitatively explain the differences, based on the new soil structure features, was recently proposed as applied to aggregated shrink-swell soils with clay content higher than the critical one, $c>c^*$ [7]. This work is a direct continuation and development of the approach as applied to aggregated shrink-swell soils with any clay content and, in particular, lower than the critical one, $c<c^*$. The basic new features of the intraaggregate structure are the rigid superficial (interface) layer of aggregates and the lacunar pores within both the layer clay and clay of the intraaggregate matrix (Fig.**2c**). Assumptions from the case $c>c^*$ [7] (Fig.**2a**) are kept in the general case (Fig.**2c**), and a new assumption about the lacunar pore volume evolution is added. In the general case the reference shrinkage curve can be predicted by eight soil parameters that can be measured independently of an observed shrinkage curve: (1) oven-dried specific volume; (2) maximum swelling water content; (3) mean solid density; (4) soil clay content; (5) oven-dried structural porosity; (6) the ratio of aggregate solid mass to solid mass of intraaggregate matrix; (7) the lacunar factor that characterizes the rate of the lacunar pore volume change with water content; and (8) oven-dried lacunar pore volume.. Two last parameters are added to the six that were used in the case $c > c^*$ [7]. The major result of the theoretical consideration and comparison between predictions and available data is as follows: existence and dewatering of a rigid superficial aggregate layer as well as existence and volume increase of intraaggregate lacunar pores at soil shrinkage permit one to quantitatively explain the origin of all the observed peculiarities of the shrinkage curve shape of aggregated soils (for enumeration of the peculiarities, see [7]). In my opinion, in spite of the limited data for checking the model, the theory and its checking are sufficiently clear and convincing, although the use of more extensive data in the future is desirable. One can expect an appreciable influence of the noted aggregate features - the interface layer with varying water content and intraaggregate lacunar pores with varying volume - not only on the soil shrinkage, but also on (i) swelling; (ii) cracking; (iii) compaction and consolidation; (iv) crusting; (v) water retention; (vi) hydraulic conductivity; and eventually (vii) water flow and solute transport.

## NOTATION

$c$     weight fraction of clay solids of the total solids (clay, silt, and sand), dimensionless

$c_v, c_w$     clay particle volume and weight fractions of all solids, dimensionless

$c^*$     critical clay content, at $c > c^*$ lacking lacunar pores, dimensionless

$F_i(w')$     saturation degree of interface pores at a given $w'$, dimensionless

$F_z$     saturation degree at the shrinkage limit of both soil intraaggregate matrix and corresponding clay matrix, dimensionless

$K$     ratio of aggregate solid mass to solid mass of intraaggregate matrix, dimensionless

$k, k_{fit}$     lacunar factor and its fitted value, dimensionless

$P_z$     oven-dried structural porosity of the soil, dimensionless

$p$     porosity of silt and sand grains coming into contact, dimensionless

$p_{av}, p_{max}, p_{min}$ average, maximum, and minimum porosity of contacting grains, dimensionless

$R_{m1}, R_{m2}$     two possible values of the maximum size of interface pores [7], μm

$R_{min}$ minimum size of interface pores [7], μm

$r_{mM}$ maximum size of clay matrix pores at the liquid limit [7], μm

$r_{mn}$ maximum size of clay matrix pores at the gravimetric water content of the intraaggregate matrix $w = w_n$ (the endpoint of the basic shrinkage of the intraaggregate matrix) [7], μm

$r_{mh}$ maximum size of clay matrix pores at the maximum swelling point, $w = w_h$ [7], μm

$S$     slope of the shrinkage curve in the basic shrinkage area, dimensionless

$U$     specific volume of soil intraaggregate matrix per unit mass of the oven-dried matrix itself; $U(w)$ is auxiliary shrinkage curve, dm$^3$ kg$^{-1}$

$U_a$     specific volume of soil aggregates (per unit mass of oven-dried soil), dm$^3$ kg$^{-1}$

$U_{az}$     oven-dried specific volume of soil aggregates, dm$^3$ kg$^{-1}$

$U_i$     contribution of interface aggregate layer to the specific volume of soil aggregates (per unit mass of oven-dried soil), dm$^3$ kg$^{-1}$

$U_s$     specific volume of structural pores (per unit mass of oven-dried soil), dm$^3$ kg$^{-1}$

$U'$     contribution of intraaggregate matrix to $U_a$ (per unit mass of oven-dried soil), dm$^3$ kg$^{-1}$

$U_z'$     oven-dried $U'$ value, dm$^3$ kg$^{-1}$

$U_{lp}$     specific volume of lacunar pores in the intraaggregate matrix, dm$^3$ kg$^{-1}$

$U_{lph}, U_{lpz}$     $U_{lp}$ value at maximum swelling and in the oven-dried state, dm$^3$ kg$^{-1}$

$u$     relative volume of intraaggregate matrix, $i.e.$, the ratio of a current soil intraaggregate matrix volume to the volume at the liquid limit, dimensionless

$u_h$     $u$ value at $w = w_h$, dimensionless

$u_h$     $u$ value at $w = w_h$, dimensionless

$u_n$     $u$ value at $w = w_n$, dimensionless

$u_S$     relative volume of nonclay solids, $i.e.$, the ratio of silt plus sand volume in the soil intraaggregate matrix to its volume at the liquid limit, dimensionless

$u_s$     relative volume of all solids, $i.e.$, the ratio of volume of all solids in the soil intraaggregate matrix to its volume at the liquid limit, dimensionless

$u_z$     $u$ value at $w = w_z$, dimensionless

$u_{cp}$     relative volume of clay pores in the intraaggregate matrix, dimensionless

$u_{lp}$     relative volume of lacunar pores in the intraaggregate matrix, dimensionless

$u_{lph}, u_{lpz}$     $u_{lp}$ value at maximum swelling and in the oven-dried state, dimensionless

$V$     current value of the specific volume of a clay matrix, dm$^3$ kg$^{-1}$

$V_z$  specific volume of clay matrix in the oven-dried syate, dm$^3$ kg$^{-1}$

$v$  relative volume of clay matrix, *i.e.*, the ratio of a current clay matrix volume to the volume at the liquid limit, dimensionless

$v_h$  relative volume of clay matrix at the maximum swelling of clay matrix, dimensionless

$v_M$=1          maximum relative volume of clay matrix at the liquid limit, dimensionless

$v_n$  relative volume of clay matrix at the air-entry point, dimensionless

$v_s$  relative volume of clay solids (clay particles) *i.e.*, the ratio of their volume to the clay matrix volume at the liquid limit [9, 10], dimensionless

$v_z$  relative volume of clay matrix at the shrinkage limit, dimensionless

$v_{cp}$  relative volume of clay pores in clay matrix, dimensionless

$v_L$  relative volume of intraaggregate clay matrix (pores and solids) at the appearance of lacunar pores at shrinkage, dimensionless

$W$  total gravimetric water content (per unit mass of oven-dried soil), kg kg$^{-1}$

$W_h$  total gravimetric water content at the maximum aggregate and soil swelling, kg kg$^{-1}$

$W_m$  maximum total gravimetric water content, kg kg$^{-1}$

$W_n$  total gravimetric water content at the endpoint of the basic shrinkage area of a soil, kg kg$^{-1}$

$W_s$  total gravimetric water content at the endpoint of the structural shrinkage area of a soil, kg kg$^{-1}$

$W_z$  total gravimetric water content at the shrinkage limit of a soil, kg kg$^{-1}$

$W_h^*$  $W_h$ plus a shear of a true saturation line relative to a pseudo one, kg kg$^{-1}$

$W_m^*$  $W_m$ plus a shear of a true saturation line relative to a pseudo one, kg kg$^{-1}$

$w$  gravimetric water content of intraaggregate matrix (per unit mass of oven-dried intraaggregate matrix itself), kg kg$^{-1}$

$w_h$  gravimetric water content of intraaggregate matrix at the maximum swelling point of the intraaggregate matrix; $w_h=W_h$, kg kg$^{-1}$

$w_M$  gravimetric water content of intraaggregate matrix at the liquid limit of the intraaggregate matrix, kg kg$^{-1}$

$w_n$  gravimetric water content of intraaggregate matrix at the endpoint of the basic shrinkage of the intraaggregate matrix, kg kg$^{-1}$

$w_z$  gravimetric water content of intraaggregate matrix at the shrinkage limit of the intraaggregate matrix, kg kg$^{-1}$

$w_L$  water content of intraaggregate matrix at the appearance of lacunar pores, kg kg$^{-1}$

$w\,'$  contribution of intraaggregate matrix to the total water content (per unit mass of oven-dried soil), kg kg$^{-1}$

$w_h'$  maximum contribution of intraaggregate matrix to the total water content at maximum aggregate swelling; $w_h'<W_h$, kg kg$^{-1}$

$w_n'$  contribution of intraaggregate matrix to the total water content at the endpoint of the basic shrinkage of the soil; $w_n'=W_n$, kg kg$^{-1}$

$w_s'$  contribution of intraaggregate matrix to the total water content at the endpoint of the structural shrinkage of the soil; $w_s'=W_s$, kg kg$^{-1}$

$w_z'$  contribution of intraaggregate matrix to the total water content at the shrinkage limit of the soil; $w_z'=W_z$, kg kg$^{-1}$

$\overline{w}$  gravimetric water content of clay matrix, kg kg$^{-1}$

$\overline{w}_h$  gravimetric water content of clay matrix at the maximum swelling point, kg kg$^{-1}$

$\overline{w}_M$  liquid limit of clay matrix, kg kg$^{-1}$

$\overline{w}_n$  gravimetric water content of clay matrix at the air-entry point, kg kg$^{-1}$

$\overline{w}_z$  shrinkage limit of clay matrix, kg kg$^{-1}$

$Y$  specific volume of a soil, dm$^3$ kg$^{-1}$

$Y_e$  experimental specific volume of a soil, dm$^3$ kg$^{-1}$

$Y_h$  $Y$ value at maximum swelling; $Y_h=Y(W_h)$, dm$^3$ kg$^{-1}$

$Y_m$  $Y$ value at $W=W_m$; $Y_m=Y(W_m)=Y_h$, dm$^3$ kg$^{-1}$

$Y_z$  oven-dried $Y$ value; $Y_z=Y(W_z)$, dm$^3$ kg$^{-1}$

$\Delta\rho$  difference between densities of clay and grain solids, g cm$^{-3}$

$\delta$  maximum relative difference between $Y$ and $Y_e$, dimensionless

$\varepsilon$  ratio of $\Delta\rho$ to $\rho_s$, dimensionless

$\zeta$  relative water content of the clay matrix or intraaggregate, dimensionless

$\zeta_L$  relative water content at the appearance of lacunar pores, dimensionless

$\zeta_M=1$     relative water content of the clay matrix or intraaggregate at the liquid limit, dimensionless

$\zeta_h$  relative water content of the clay matrix or intraaggregate at maximum swelling, dimensionless

$\zeta_n$  relative water content of the clay matrix or intraaggregate at the air-entry point, dimensionless

$\zeta_z$  relative water content of the clay matrix or intraaggregate at the shrinkage limit, dimensionless

$\Pi$  porosity of interface aggregate layer, dimensionless

$\rho_b$  dry aggregate bulk density, g cm$^{-3}$

$\rho_{cl}$  density of clay solids (clay particles), g cm$^{-3}$

$\rho_s$  mean density of solids, g cm$^{-3}$

$\rho_{ss}$  density of silt and sand grains, g cm$^{-3}$

$\rho_w$  density of water, g cm$^{-3}$

**Figure captions**

**Fig.1**. The qualitative look of (a) the clay and (b) soil shrinkage curves (for the designations see Notation). The curve slope in the basic shrinkage area for soil ($W_n<W<W_s$) can be less than unity, unlike for clay ($\overline{w}_n < \overline{w} < \overline{w}_h$). Curves 1 and 2 are the observed variants at structural shrinkage. Dashed and dash-dot lines are the true and pseudo saturation lines, respectively.

**Fig.2**. The illustrative scheme of the internal structure of aggregates at a clay content: (a) $c>c^*$, without lacunar pores; (b) $c<c^*$, with lacunar pores and silt-sand grain contacts at $w=0$; and (c) $c<c^*$, with lacunar pores and non-totally contacting silt-sand grains at any water content.

**Fig.3**. Variants of the shrinkage curve $U(w)$ of the intraaggregate matrix (corresponding to $V(\overline{w})$ of contributing clay in Fig.**1a**; with $w=\overline{w} c$) depending on the clay content (for the designations see Notation). (a) $c>c^*$; lacunar pores are lacking at $0<w\leq w_h$ (Fig.**2a**); $k=0$; at $w_n<w<w_h$ the slope $dU/dw=1/\rho_w$; $U(w)$ starts at the true saturation line (dashed line). (b) $c<c^*$; lacunar pores appear at $w=w_L<w_h$ (Fig.**2c**); at $w_L<w\leq w_h$ $k=0$ and the slope $dU/dw=1/\rho_w$; at $w_n\leq w\leq w_L$ $0<k=$const$\leq 1$ and the slope $dU/dw=(1-k)/\rho_w<1/\rho_w$; $U(w)$ starts at the true saturation line. (c) $c<c^*$; lacunar pores exist at $0<w\leq w_h$ (Fig.**2c**); $0<k=$const$\leq 1$; at $w_n<w<w_h$ the slope $dU/dw=(1-k)/\rho_w<1/\rho_w$; $U(w)$ starts at the pseudo saturation (dash-dot) line (with that $w_h^*$-$w_h=U_{lp}(w_h)\rho_w$).

**Fig.4**. A qualitative view of the soil reference shrinkage curve, $Y(W)$ and that of aggregates, $U_a(W)$ (for options 1 and 2) when the clay content $c<c^*$ as well as a number of auxiliary (dotted) curves: $Y(w')$, $U_a(w')$, $U(w)$, and $U'(w')$ (for the designations see Notation). Dashed and dash-dot inclined straight lines are the true and pseudo saturation lines, respectively, for both the $w$ and $W$ axes.

**Fig.5**. Shrinkage data (white squares) and prediction (solid line of type 1) for ferruginous soil horizon A from [11] (for the designations see Fig.4 and Notation). The dotted line is parallel to the shrinkage curve in the basic shrinkage area.

**Fig.6.** As in Fig.5 for ferruginous soil horizon B1 from [11]. The solid line is that of type 2.

**Fig.7.** As in Fig.5 for ferruginous soil horizon B2 from [11]. The solid line is that of type 2.

**Fig.8. As** in Fig.5 for ferruginous soil horizon AB from [11]. The solid line is that of type 1.

**Fig.9. As** in Fig.5 for ferralitic soil horizon A from [11]. The solid line is that of type 2.

**Fig.10. As** in Fig.5 for ferralitic soil horizon B1 from [11]. The solid line is that of type 2.

**Fig.11. As** in Fig.5 for ferralitic soil horizon B2 from [11]. The solid line is that of type 2.

Fig.1

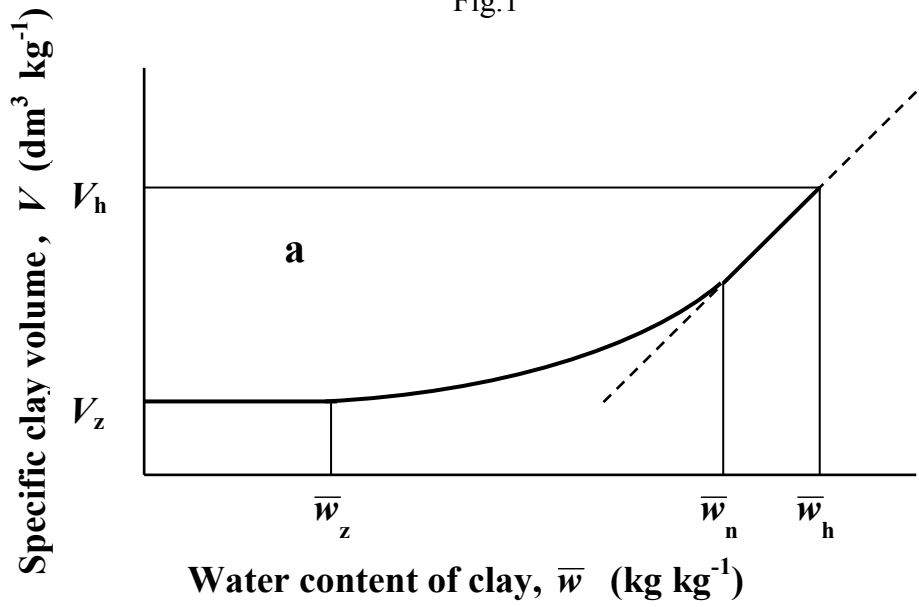

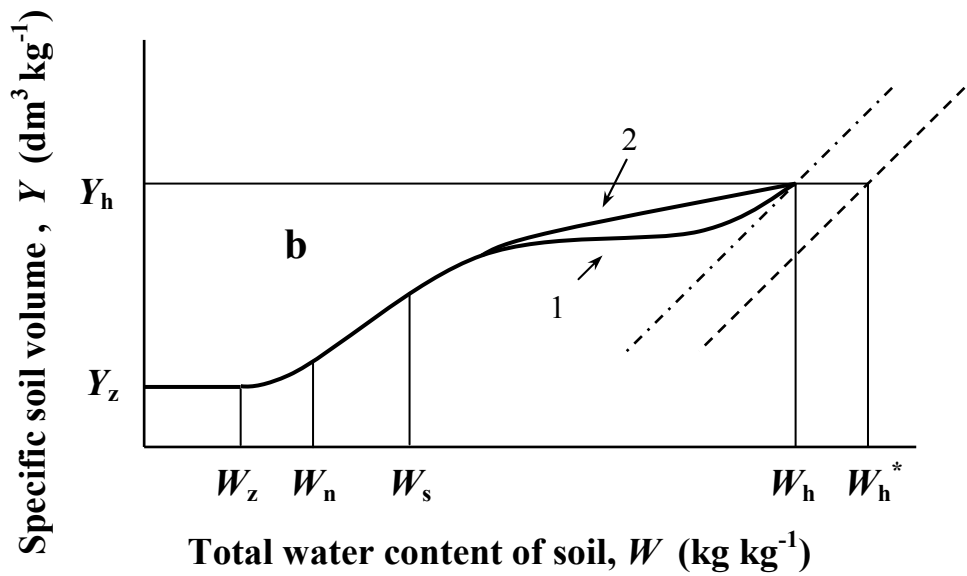

Fig.2

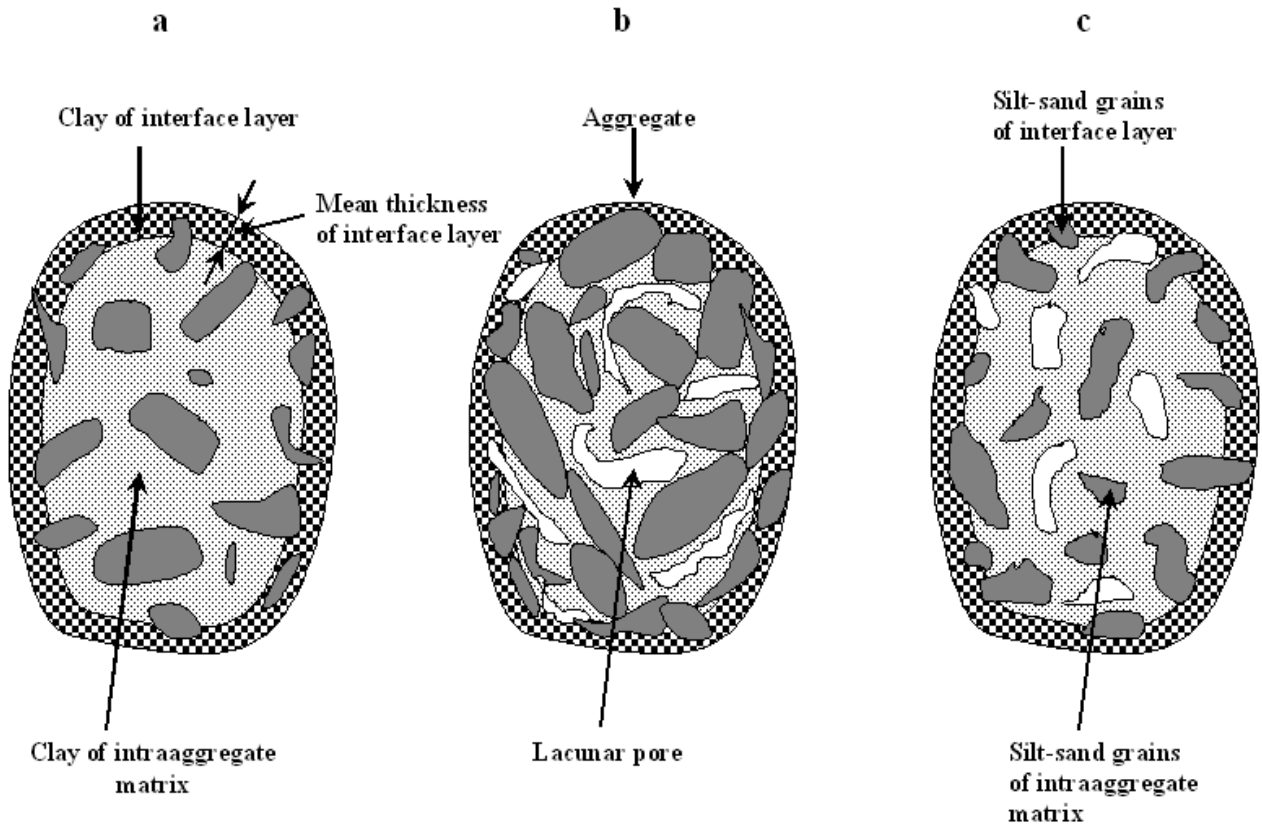

a            b            c

Clay of interface layer

Mean thickness
of interface layer

Aggregate

Silt-sand grains
of interface layer

Clay of intraaggregate
matrix

Lacunar pore

Silt-sand grains
of intraaggregate
matrix

Fig.3

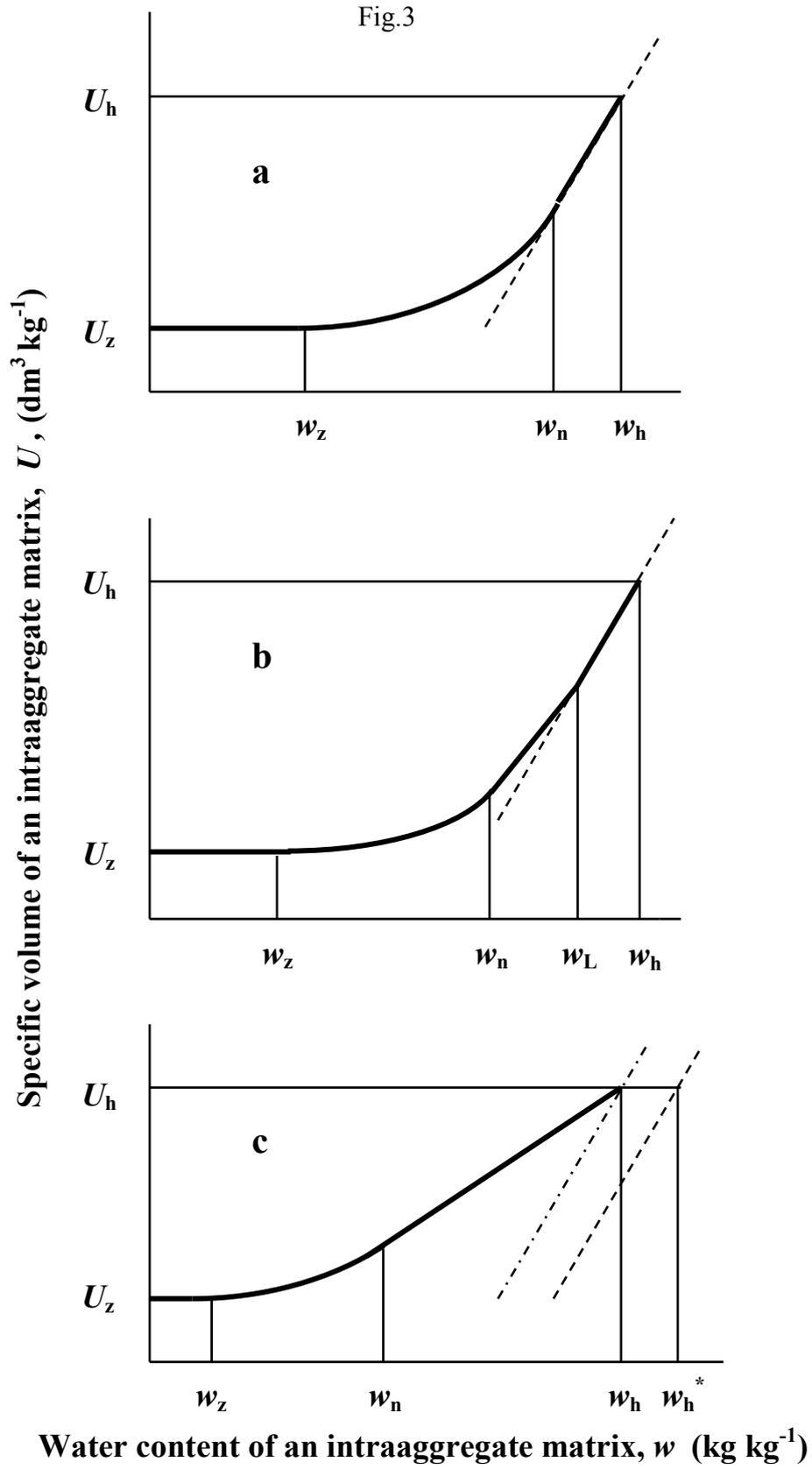

**Water content of an intraaggregate matrix, *w* (kg kg⁻¹)**

Fig.4

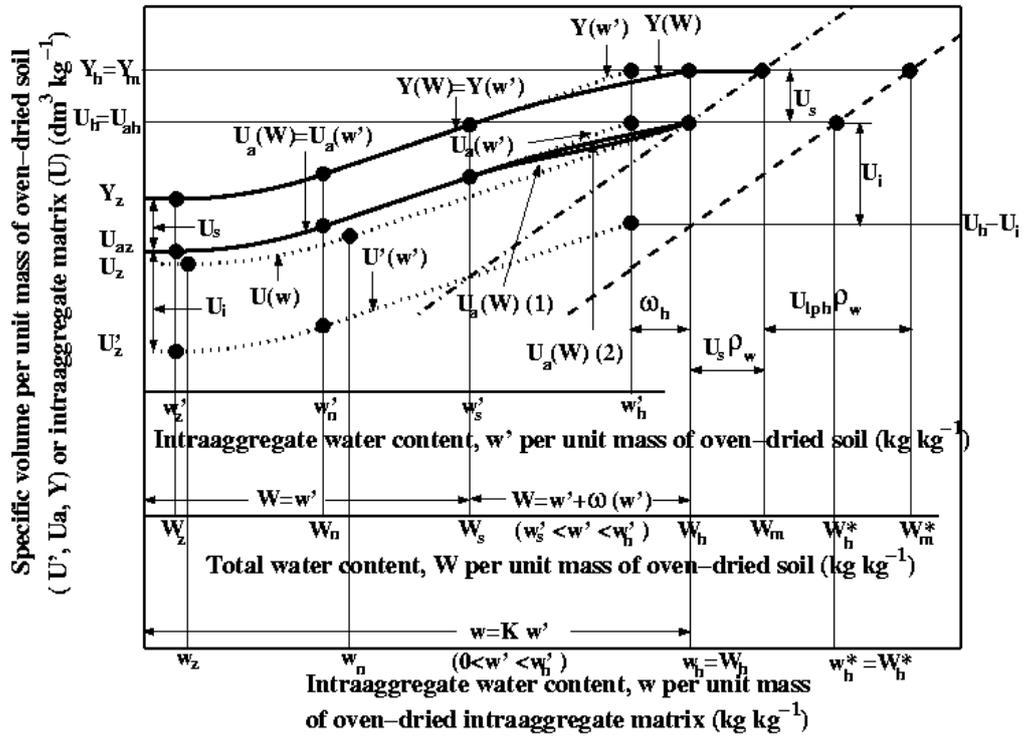

Fig.5

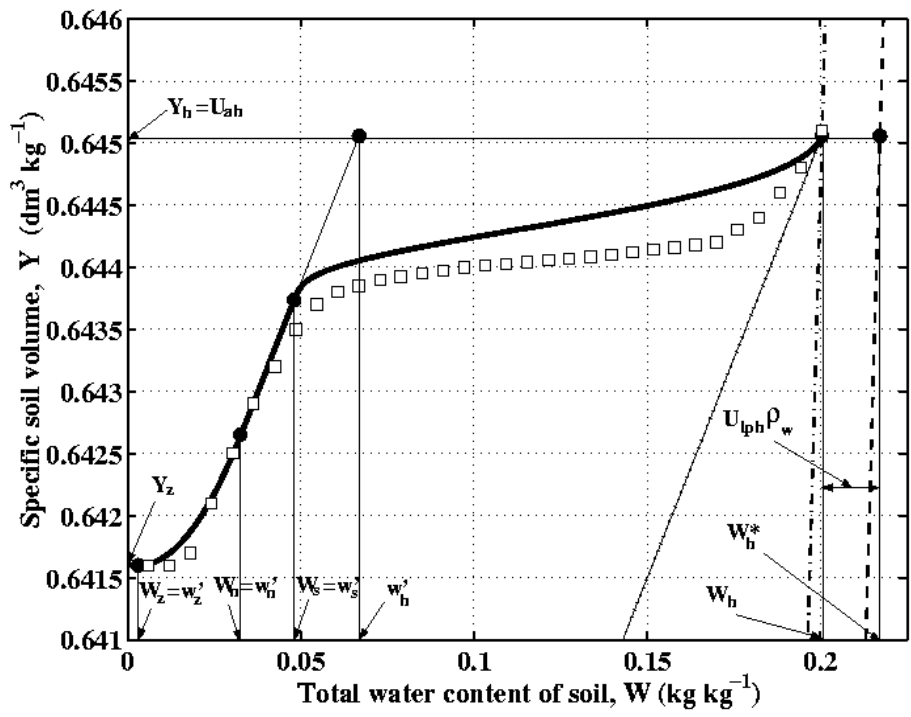

Fig.6

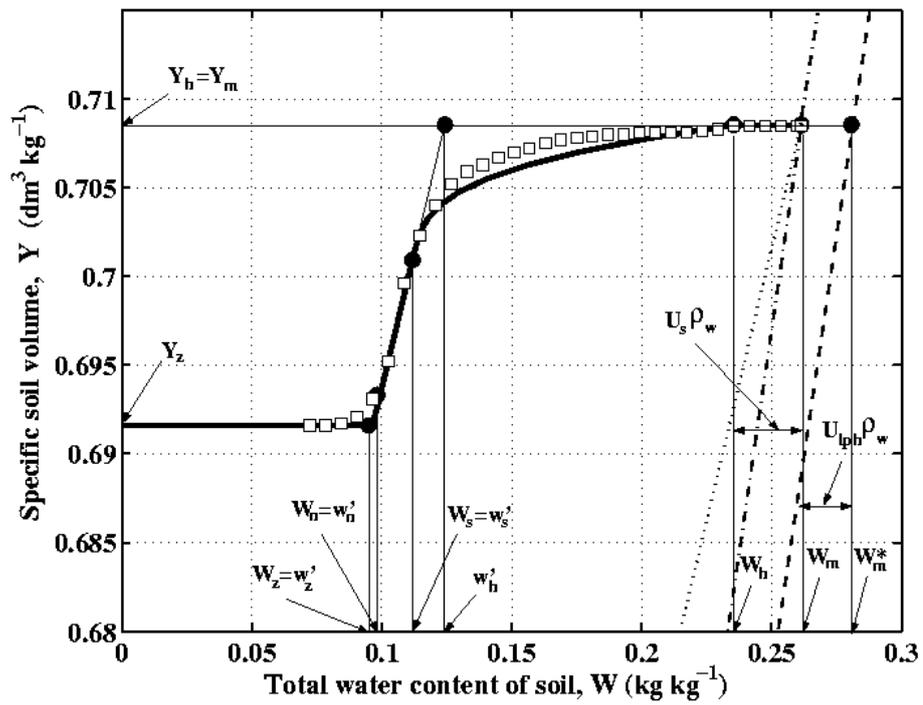

Fig.7

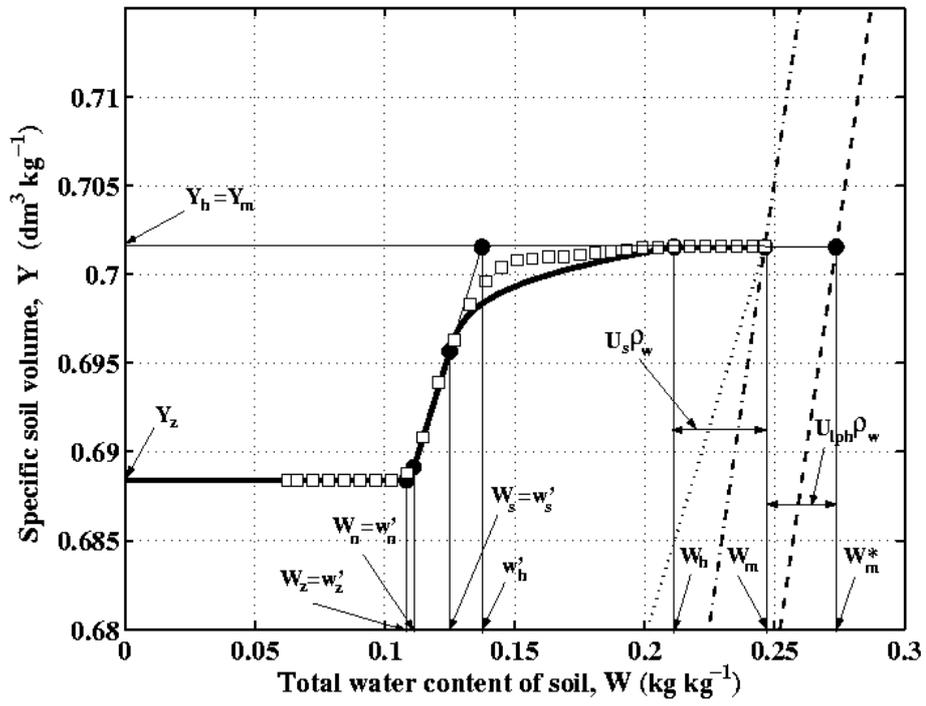

Fig.8

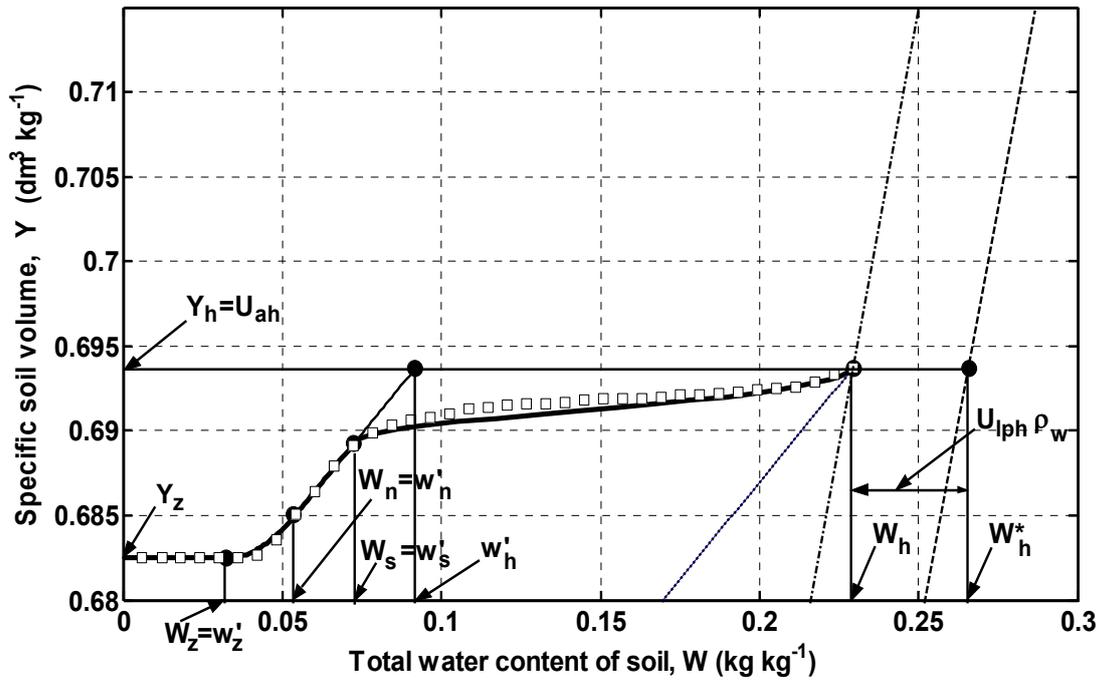

Fig.9

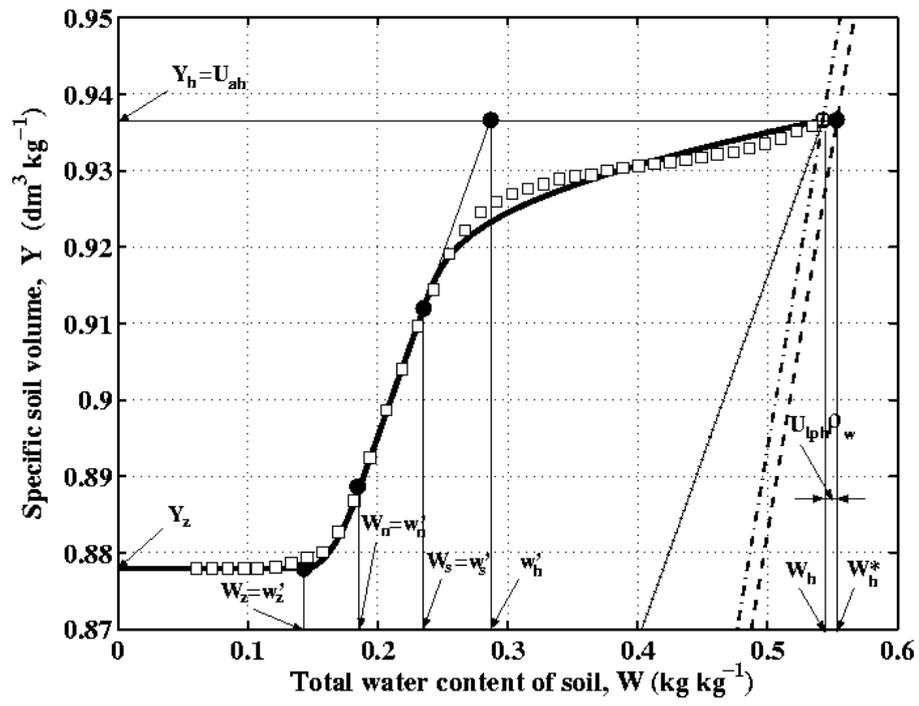

Fig.10

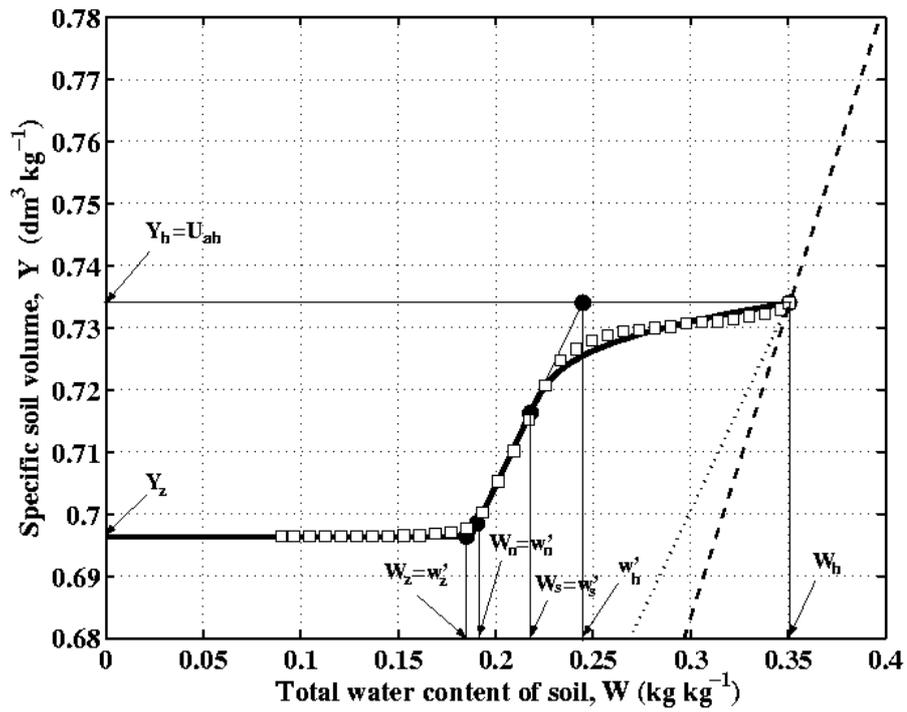

Fig.11

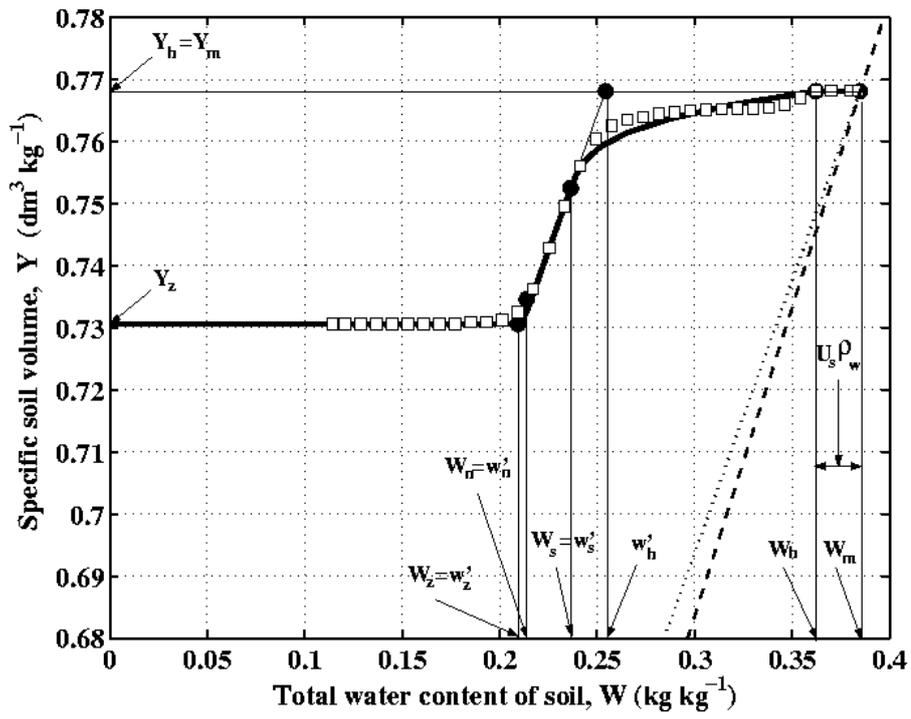

Table 1. Input parameters of the seven soils for reference shrinkage curve prediction, maximum relative difference (δ) between predicted and measured volume shrinkage, and parameters for comparison with fitted and predicted ones.

| Data source | | Fig | Input parameters[†] | | | | | | | | | Parameters for comparison[‡] | | | | |
|---|---|---|---|---|---|---|---|---|---|---|---|---|---|---|---|---|
| | | | $Y_z$ | $W_h$ | $\rho_s$ | $c$ | $P_z$ | $K_{fit}$ | $k_{fit}$ | $U_{lpzfit}$ | $\delta$ | $W_h/W_h'$ | $S$ | $W_m$ | $W_h'$ | $W_m'$ |
| | | | $dm^3\,kg^{-1}$ | $kg\,kg^{-1}$ | $g\,cm^{-3}$ | | | | | $dm^3\,kg^{-1}$ | % | | | $kg\,kg^{-1}$ | | |
| Ferruginous soil from Fig.3 of [11] | A horizon | 5 | 0.642 | 0.200 | 2.337 | 0.065 | 0 | 3 | 0.929 | 0.153 | 0.06 | 3 | 0.071 | $=W_h$ | 0.217 | $=W_h^*$ |
| | B1 horizon | 6 | 0.692 | 0.235 | 2.337 | 0.340 | 0.038 | 1.894 | 0.387 | 0.040 | 0.12 | 1.894 | 0.613 | 0.261 | 0.255 | 0.281 |
| | B2 horizon | 7 | 0.688 | 0.211 | 2.337 | 0.353 | 0.051 | 1.538 | 0.524 | 0.050 | 0.20 | 1.538 | 0.476 | 0.246 | 0.239 | 0.274 |
| | AB horizon | 8 | 0.682 | 0.229 | 2.337 | 0.187 | 0 | 2.499 | 0.771 | 0.130 | 0.08 | 2.499 | 0.229 | $=W_h$ | 0.266 | $=W_h^*$ |
| Ferralitic soil from Fig.3 of [11] | A horizon | 9 | 0.878 | 0.543 | 2.608 | 0.523 | 0 | 1.892 | 0.523 | 0.132 | 0.22 | 1.892 | 0.477 | $=W_h$ | 0.553 | $=W_h^*$ |
| | B1 horizon | 10 | 0.696 | 0.351 | 2.608 | 0.648 | 0 | 1.433 | 0.340 | 0.028 | 0.23 | 1.433 | 0.660 | $=W_h$ | $=W_h$ | $=W_h$ |
| | B2 horizon | 11 | 0.731 | 0.362 | 2.608 | 0.648 | 0.031 | 1.424 | 0.121 | 0.0074 | 0.33 | 1.424 | 0.879 | 0.385 | $=W_h$ | $=W_m$ |

[†] $Y_z$, specific volume of an oven-dried soil; $W_h$, total gravimetric water content at maximum aggregate and soil swelling; $\rho_s$, mean density of solids; $c$, weight fraction of clay solids; $P_z$, oven-dried structural porosity of the soil; $K_{fit}$, fitted aggregate/intraaggregate mass ratio; $k_{fit}$, fitted lacunar factor value; $U_{lpzfit}$, fitted oven-dried value of the specific volume of lacunar pores.

[‡] $W_h/W_h'$, experimental estimate of the $K$ ratio; $S$, experimental estimate of the shrinkage curve slope in the basic shrinkage area; $W_m$, experimental estimate of the maximum total water content of a soil; $W_h^*$, the experimental $W_h$ value plus a shear of true saturation line relative to pseudo one; $W_m^*$, the experimental $W_m$ value plus a shear of the true saturation line relative to the pseudo one.

**Table 2.** Predicted parameters of the seven soils.

| Fig | Parameters of intraaggregate matrix[†] | | | | | Lacunar pore parameters[‡] | | Parameters of interface layer[§] | | | | | Parameters of soil as a whole[¶] | | | |
|---|---|---|---|---|---|---|---|---|---|---|---|---|---|---|---|---|
| | $u_s$ | $u_z$ | $u_S$ | $w_z$ | $w_n$ | $U_{lpm}$ | $U_{lpb}$ | $U_l$ | $\Pi$ | $R_{min}/r_{mM}$ | $R_{m1}/r_{mM}$ | $R_{m2}/r_{mM}$ | $U_s$ | $W_z$ | $W_n$ | $W_s$ |
| | | | — kg kg⁻¹ — | | | — dm³ kg⁻¹ — | | | | | | | dm³ kg⁻¹ | — kg kg⁻¹ — | | |
| 5 | 0.516 | 0.766 | 0.483 | 0.009 | 0.098 | 0.112 | 0.017 | 0.430 | 0.311 | 0.729 | 0.803 | 0.902 | 0 | 0.003 | 0.033 | 0.048 |
| 6 | 0.476 | 0.724 | 0.314 | 0.180 | 0.186 | 0.038 | 0.019 | 0.322 | 0.345 | 0.808 | 0.828 | 0.914 | 0.026 | 0.095 | 0.098 | 0.112 |
| 7 | 0.503 | 0.760 | 0.326 | 0.167 | 0.171 | 0.049 | 0.027 | 0.233 | 0.317 | 0.814 | 0.832 | 0.916 | 0.035 | 0.108 | 0.111 | 0.125 |
| 8 | 0.483 | 0.751 | 0.392 | 0.081 | 0.134 | 0.110 | 0.036 | 0.416 | 0.331 | 0.766 | 0.815 | 0.908 | 0 | 0.032 | 0.054 | 0.072 |
| 9 | 0.261 | 0.562 | 0.124 | 0.271 | 0.349 | 0.112 | 0.010 | 0.442 | 0.579 | 0.776 | 0.816 | 0.908 | 0 | 0.143 | 0.185 | 0.235 |
| 10 | 0.354 | 0.627 | 0.124 | 0.265 | 0.273 | 0.026 | 0 | 0.222 | 0.478 | 0.811 | 0.831 | 0.916 | 0 | 0.185 | 0.190 | 0.218 |
| 11 | 0.346 | 0.625 | 0.122 | 0.298 | 0.304 | 0.007 | 0 | 0.222 | 0.486 | 0.817 | 0.830 | 0.915 | 0.022 | 0.209 | 0.214 | 0.236 |

[†] $u_s$, relative volume of all solids; $u_z$, relative volume of intraaggregate matrix at its shrinkage limit; $u_S$, relative volume of nonclay solids; $w_z$, gravimetric water content of intraaggregate matrix at the shrinkage limit of the intraaggregate matrix; $w_n$, gravimetric water content of intraaggregate matrix at the endpoint of the basic shrinkage of the intraaggregate matrix.

[‡] $U_{lpm}$, specific volume of lacunar pores within the intraaggregate matrix at the endpoint of its basic shrinkage; $U_{lpb}$, specific volume of lacunar pores within the intraaggregate matrix at the maximum swelling point.

[§] $U_l$, contribution of the interface aggregate layer to the specific volume of soil aggregates; $\Pi$, porosity of the interface layer; $R_{min}$, minimum size of interface pores [7]; $R_{m1}$ and $R_{m2}$, two possible values of the maximum size of interface pores [7]; $r_{mM}$, maximum size of clay matrix pores at the liquid limit [7].

[¶] $U_s$, specific volume of structural pores; $W_z$, total gravimetric water content at the shrinkage limit; $W_n$, total gravimetric water content at the endpoint of the basic shrinkage area; $W_s$, total gravimetric water content at the endpoint of the structural shrinkage area.

**Table 3.** Predicted parameters of the clays contributing to the seven soils, porosity of contacting grains, and critical soil clay content ($c^*$).

| Fig | Relative clay macroparameters[†] | | | | | Relative clay microparameters[‡] | | Porosity of contacting grains[§] | | | $c^*(p_{av})$ |
|---|---|---|---|---|---|---|---|---|---|---|---|
| | $v_s$ | $v_z$ | $v_h$ | $v_L$ | $F_z$ | $r_{mr}/r_{mM}$ | $r_{mb}/r_{mM}$ | $p_{min}$ | $p_{max}$ | $p_{av}$ | |
| 5 | 0.065 | 0.190 | 0.532 | 0.574 | 0.172 | 0.654 | 0.803 | 0.169 | 0.370 | 0.269 | 0.112 |
| 6 | 0.236 | 0.533 | 0.618 | 0.699 | 0.983 | 0.789 | 0.828 | 0.538 | 0.566 | 0.552 | 0.353 |
| 7 | 0.263 | 0.558 | 0.632 | 0.723 | 0.988 | 0.797 | 0.832 | 0.536 | 0.572 | 0.554 | 0.370 |
| 8 | 0.148 | 0.348 | 0.574 | 0.662 | 0.755 | 0.716 | 0.815 | 0.350 | 0.477 | 0.414 | 0.232 |
| 9 | 0.156 | 0.397 | 0.578 | 0.594 | 0.874 | 0.735 | 0.816 | 0.736 | 0.778 | 0.757 | 0.551 |
| 10 | 0.262 | 0.545 | 0.631 | $=v_h$ | 0.985 | 0.790 | 0.831 | 0.793 | 0.802 | 0.797 | 0.654 |
| 11 | 0.255 | 0.565 | 0.628 | $=v_h$ | 0.989 | 0.803 | 0.830 | 0.803 | 0.805 | 0.804 | 0.650 |

[†] $v_s$, relative volume of clay solids at the liquid limit; $v_z$, relative volume of clay matrix in the oven-dried state; $v_h$, relative volume of clay matrix at the maximum swelling; $v_L$, relative volume of intraaggregate clay matrix at the appearance of lacunar pores at soil shrinkage; $F_z$, saturation degree of clay matrix at the shrinkage limit.

[‡] $r_{mM}$, maximum size of clay matrix pores when the gravimetric water content of the intraaggregate matrix is at the endpoint of the basic shrinkage of the intraaggregate matrix [7]; $r_{mb}$, maximum size of clay matrix pores at the maximum swelling point [7]; $r_{mb}$, maximum size of clay matrix pores at the liquid limit [7].

[§] $p_{min}$, $p_{max}$, and $p_{av}$, minimum, maximum, and average porosity of contacting grains, respectively.